\documentclass[dvips,6pt,cmp,aps,twocolumn]{revtex4}

\usepackage{amsmath,amssymb,epsf,epsfig,amsthm,bm,graphicx,subfigure,natbib}

\newcommand{\cL}{{\cal L}}

\newcommand{\cR}{{\cal R}}

\newcommand{\bcR}{\partial{\cal R}}

\newcommand{\bda}{\boldsymbol{\delta}}
\newcommand{\bds}{\boldsymbol{\sigma}}

\theoremstyle{theorem}
\newtheorem{theorem}{Theorem}[section]

\theoremstyle{definition}

\newtheorem{definition}[theorem]{Definition}

\newtheorem{remark}[theorem]{Remark}

\newtheorem{note}[theorem]{Note}

\begin{document}

\title{Conserved charges in (Lovelock) gravity in first order formalism}%

\author{Elias Gravanis }
\email{eliasgravanis@netscape.net}
\date{\today}%

\begin{abstract}

We derive conserved charges as quasi-local Hamiltonians by covariant
phase space methods for a class of geometric Lagrangians that can be
written in terms of the spin connection, the vielbein and possibly
other tensorial form fields, allowing also for non-zero torsion. We
then re-calculate certain known results and derive some new ones in
three to six dimensions hopefully enlightening certain aspects of
all of them. The quasi-local energy is defined in terms of the
metric and not its first derivatives, requiring `regularization' for
convergence in most cases. Counter-terms consistent with Dirichlet
boundary conditions in first order formalism are shown to be an
efficient way to remove divergencies and derive the values of
conserved charges, the clear-cut application being metrics with AdS
(or dS) asymptotics. The emerging scheme is: all is required to
remove the divergencies of a Lovelock gravity is a boundary Lovelock
gravity.

\end{abstract}

\maketitle


\setcounter{footnote}{0}
\section{Introduction}

Defining energy in a consistent and general manner is exceedingly
difficult in general relativity. One good reason is that unlike any
other field theory it is hard to localize energy in gravity: its
density should be reasonably constructed out of first derivatives of
the metric which vanish at any point for inertial observers.
Formally, energy is the generator of a local symmetry thus it should
vanish locally. Another reason of similar nature, is that energy, by
shaping spacetime itself, is not defined in some absolute manner: it
depends strongly on boundary conditions, which introduce the
asymptotics i.e the class of metrics one intends to study.
Implicitly or explicitly the conditions define a reference
background, for which energy is zero.

In general relativity, the Bondi-Sachs
mass~\cite{Bondi:1962px}\cite{Sachs:1962wk} defined in the null
infinity and the ADM mass~\cite{Arnowitt:1959ah} defined in the
spatial infinity, are accepted definitions of energy for
asymptotically flat spacetimes. ADM mass is interpreted as the total
energy of spacetime. Abbott-Deser mass~\cite{Abbott:1981ff} provides
a definition for asymptotically de Sitter spacetimes, in the spirit
of ADM mass. All these definitions are written as an integral over a
co-dimension two spatial sphere. This is an implication of
non-localizability.

Quasi-localization is a most interesting alternative. The word
means: we consider the total energy content of finite spatial
regions, of size (and shape) otherwise arbitrary. Of all
definitions, some of which are mentioned in section
\ref{conservation}, the Brown-York quasi-local
energy~\cite{Brown:1992br} is especially appealing. By the
generalization of
\cite{Balasubramanian:1999re}\cite{Balasubramanian:2001nb}, which
originated from the `\textit{AdS}/CFT
correspondence'~\cite{Maldacena:1997re}\cite{Witten:1998qj}, the
definition applies to spacetimes with \textit{AdS} or even de Sitter
asymptotics in a natural manner. Moreover, both the Brown-York
definition and the famous correspondence are in the spirit of
`holography'~\cite{'tHooft:1993gx}\cite{Susskind:1994vu}: the
gravitational degrees of freedom are fundamentally boundary degrees
of freedom.

Even if one is not particularly interested in the localization
problem, the Brown-York quasi-local energy, as a means to define
energy in spacetime, is technically attractive. It derives from the
invariant action functional as a generator in the sense of
Hamilton-Jacobi theory~\cite{Brown:1992br}\cite{Brown:2000dz}. Such
a definition is still applicable when one departs from general
relativity.

It has become customary in recent years to consider curved
spacetimes of dimension higher than four. Once such a choice is made
the elementary conditions that uniquely determine the Einstein
equations in four dimensions are not adequate. In dimension five or
higher there exist rank two, symmetric and covariantly conserved
tensors involving up to second derivatives of the metric tensor.
These are the Lovelock tensors~\cite{Lovelock:1971yv}. Generic
linear combinations of them define equally good field equations for
the metric tensor. Moveover the field equations are second order
differential equations, giving `Lovelock gravity' a feeling of
familiar ground. The Cauchy problem appears almost- though not
entirely- analogous to that in general
relativity~\cite{ChoquetBruhat:1988dw}.

The action functional for Lovelock gravity is very interesting
(starting from the Einstein-Hilbert action itself). Written in
tensor notation, each Lovelock Lagrangian has the form of a
topological density of some lower (even) dimension, only now its
indices are summed over a higher number of dimensions. This
dimensional translation provides \emph{local} dynamics for the
metric tensor: Einstein and general Lovelock field equations.

This is mysteriously neat. The mystery is partly dissolved if we, as
Zumino did~\cite{Zumino:1985dp}, get more geometric. For our needs
useful mathematical references are
\cite{Eguchi:1980jx}\cite{goldberg}. The metric tensor may be
replaced by the vielbein $E^a$ i.e. a basis in an abstract Lorentz
vector bundle with metric $\eta_{ab}$ and volume form $\epsilon_{a
\cdots b}$. The spacetime metric tensor reads $g=\eta_{ab} E^a
\otimes E^b$. A connection $\omega^a_{~b}$ on the Lorentz bundle may
or may not be Levi-Civita, and let
$\Omega^a_{~b}:=d\omega^a_{~b}+\omega^a_{~c}\omega^c_{~b}$ denote
its curvature. (The exterior product is understood). The general
Lovelock Lagrangian is constructed by the Lorentz invariant form
$\epsilon(\Omega \cdots \Omega E \cdots E) \equiv \epsilon_{ab
\cdots cde \cdots f} \Omega^{ab} \cdots \Omega^{cd} E^e \cdots E^f$.
We integrate this density over a manifold without boundary, to get
an action. The difference between Einstein and Lovelock gravity is
that the latter contains terms with more than one $\Omega$ factors.
First order formalism means treating $E$ and $\omega$ as independent
variables.

Varying w.r.t. $\omega$ the curvature changes by $\delta
\Omega=D(\delta \omega)$, where $D$ is the covariant derivative. The
curvature satisfies $D\Omega=0$, the Bianchi identity. Varying the
action w.r.t. to $\omega$ we get, after an integration by parts,
terms involving $DE$. If there no $E$ factors in the Lagrangian the
variation vanishes identically: it is a topological density. The
torsion, $DE$, vanishes iff the connection is Levi-Civita. Upon
restricting ourselves to such connections the variation vanishes
identically (in Einstein gravity we have no choice); then the
Lovelock tensors and field equations are obtained, purely
algebraically, by varying the action w.r.t. $E$.

Technically there is a bonus in the first order formulation. Once
one is free to allow for torsion, one may consider a generalization
of Lovelock gravity such that connection is not Levi-Civita, as well
as consider other Lorentz invariant densities built by contraction
with the invariant tensors $\delta^a_{[b} \cdots \delta^d_{c]}$.
These are related to the Pontryagin invariants as opposed to
Lovelock terms which are related to the Euler number. These Lorentz
invariants will be our toy model of a geometric gravitational
Lagrangian. Most of our examples though will be in Lovelock gravity.

In section \ref{action and symplectic form} the covariant phase
space methods as applied to a gravitational action are discussed. In
section \ref{section-type-I} we derive the Hamiltonian in this
framework based on the Poincare invariance of the symplectic form,
and distinction between cases where this can and cannot be done is
made, type and $(\textbf{I})$ and $(\textbf{II})$ cases
respectively. In section \ref{conservation} the quasi-local value of
the charges is introduced. In section \ref{The Hamiltonian generator
in type II} the type $(\textbf{II})$ case Hamiltonian is derived.
Between sections \ref{Spherically symmetric metrics}-\ref{Five
dimensional Chern-Simons point particle} various results in Einstein
gravity including torsion are collected and discussed. Matters such
as a possibility for attributing a non-zero mass to Minkowski space
in certain gravity theories is analyzed. In section
\ref{Boulware-Deser-Cai metrics} an example of Lovelock gravity, the
case of five dimensions, is discussed in detail. In section
\ref{Boulware-Deser metrics in $6d$} a brief excursion in dimension
six shows that the Dirichlet boundary forms are adequate to remove
the divergencies and obtain results obtained by other methods. It is
also understood that there is a general scheme: at least in the
first order formalism, all is required to remove the divergencies of
a Lovelock gravity is a boundary Lovelock gravity. In section
\ref{interpolations} the quasi-local mass is used to interpolate
between \textit{AdS} and flat spacetime results. A brief discussion
on boundary forms is in the appendix.

As a matter of terminology, one should distinguish the `canonical
generator' in the sense of Hamiltonian theory which involves Poisson
brackets, from that in the sense of Hamilton-Jacobi theory where
generators are defined as derivatives of Hamilton's Principal
Function that is the action functional on-shell. Of course, the two
kinds are intimately related; the covariant phase space framework
provides the link. But there are differences. One obtains
respectively: the value of the total (over all space) value of the
charge, and the actual quasi-local value over a finite spatial
region. But we shall work everything out in a quasi-local setting:
the boundary Lagrangian will be of crucial importance. Thus
everything is quasi-local and what is what will be understood from
context. After all, in the covariant phase formulation one can
easily see that the family of functions (with radius as parameter)
whose limit is the Hamiltonian generator are the Hamilton-Jacobi
generators, section \ref{conservation}. Having said that we shall
not be very careful about terminology for the rest of this paper.

\section{The conserved charge as canonical
generator}\label{basic_formula}

\subsection{Action and invariant symplectic form}
\label{action and symplectic form}

Let  a region ${\cal R}$ be an arbitrary open set spacetime $M$. Let
$\Psi^{A\dots}$ be a collection of vector valued forms over $M$.
(The Lorentz indices of the bulk fields are now denoted by capital
letters.)Let fields intrinsic to the boundary $\partial { \cal R}$
of the region ${\cal R}$ be denoted by $\Psi^{a \dots}_{\|}$. We
will use this same symbol $\Psi_{\|}$ for bulk fields which agree
with the intrinsic ones when pulled back into $\cR$.

The induced fields into the boundary are $i^*_{\|} \Psi^{a\dots}$.
In every smooth component of the boundary $\bcR$, and as long as
these components are `orthogonal'~\cite{Brown:2000dz}, the intrinsic
fields can be chosen to coincide with the induced values of the bulk
fields: $\Psi^{a\dots}_{\|}=i^*_{\|} \Psi^{a\dots}$. We shall be
content with the `orthogonal' case.

Consider a local functional of the fields
\begin{equation}\label{action_in_R}
S=\int_{\cR} {\cal L}+ \int_{\bcR} i^*_{\|}{\cal B}_{(C)}\,,
\end{equation}
where ${\cal L}$ is a functional of the fields $\Psi$, the
Lagrangian of the theory. ${\cal B}_{(C)}$ is a form in $\cR$ that
depends on $\Psi$ and $\Psi_{\|}$ and it is such that: Under given
conditions $C$ for the variations $\delta_{(C)}$ of the fields
$\Psi$ and $\Psi_{\|}$, the action functional $S$ has an extremum
\begin{equation}\label{extremized}
\delta_{(C)}S=0\,,
\end{equation}
when the Euler-Lagrange equations are satisfied by $\Psi$ in $\cR$.

We want first to find and analyze a formula for the total energy and
angular momentum, derived as canonical generators conjugate to the
associated diffeomorphisms, for the theory of fields $\Psi$ given by
the Lagrangian ${\cal L}$. Our variables $\Psi$ will be the vielbein
$E$ and the Lorentz connection $\omega$, as well as other `matter'
fields which we shall not explicitly mention. Our toy model, and
quite the objective, will be Lagrangian ${\cal L}$ which is a linear
combination of local Lorentz invariants of the curvature $\Omega$
and $E$. There can also be `matter' Lagrangian terms which will be
left understood.

As a warmup, consider an action in particle mechanics:
$S=\int_{t_1}^{t_2} dt\, L(q,\dot q)$. Let's define the phase space
of the theory to mean the space of solutions of the classical
equations~\cite{Crnkovic:1986ex}, denoted compactly as
$\mathcal{E}=0$. On the phase space of the histories $q$
infinitesimal variations $\delta q$ are tangent directions on that
space and can be thought of as vector fields. They are solutions to
the linearized equations of motion $\delta \mathcal{E}=0$. It is
quite convenient to define an exterior derivative $\bda$ over the
phase space. Its nilpotence, $\bda^2=0$, acting on the fields means
that: We choose a multi-parameter family of fields twice
continuously differentiable w.r.t. these parameters, variation means
differentiation w.r.t. a parameter and the order of differentiations
is irrelevant. We have:
\begin{equation}\label{}
\bda^2 S=\left[ \bda p\, \bda q \right]_{t_1}^{t_2}-
\int_{t_1}^{t_2} dt\, \bda q \cdot \bda \mathcal{E}\,,
\end{equation}
where of course $p=\partial L/\partial \dot q$. By the nilpotence of
$\bda$ we have that $\bda^2 S=0$. So for on-shell variations we get
$\left[ \bda p\, \bda q \right]_{t_1}^{t_2}=0$ i.e. $\bda p\, \bda
q$ is unchanged through time.

The importance of the phase space 2-form $\bds:= \bda p\, \bda q$ is
that it gives the clue to formulate the Hamiltonian theory in a
relativistically covariant way~\cite{Crnkovic:1986ex}. Phase space
has here explicit coordinates $x=(x^1,x^2)=(p,q)$. The form $\bds$
is closed $\bda \bds=0$ and non-degenerate i.e. its components
$\bds_{12}=-\bds_{21}=1$ form an invertible matrix. Now if on a
phase space we can find an closed and non-degenerate two-form we say
we have a \textit{symplectic structure}. The crucial theorem of
Darboux~\cite{arnold} says that we can always find coordinates such
that locally a symplectic structure can be written in the form $\bda
p\, \bda q$. Therefore we have the essential structure to do
canonical theory without any unnessecary choice of coordinates that
breaks covariance. If the symplectic structure does not depend on
time, as above, then it is a Poincare invariant. Thus one succeeds
in formulating a covariant canonical theory. Hamilton's equations
can be written as $\dot x^i \bds_{ij}=-\partial_j H$. Phase space
components need not necessarily appear as we will see below. The
symplectic methods in Hamiltonian and Lagrangian mechanics are
discussed in depth in \cite{marsden} and \cite{arnold}.

Let's turn now to the theory of fields $\Psi$ and our toy model. We
have
\begin{equation}\label{first-variation}
\bda S=\int_{\bcR} \bda \omega \cdot \frac{\partial \cL}{\partial
\Omega}+\bda {\cal B}_{(C)}+\int_{\cR} \bda\Psi \cdot \mathcal{E}\,.
\end{equation}
$\bda$ is thought of as anti-commuting with the exterior derivative
$d$, $\bda d+d \bda=0$, and the same with all $p$-forms with $p$
odd.

Taking the second variation we have:
\begin{equation}\label{second-variation}
0=\bda^2 S=\int_{\bcR} \bda\Big\{\bda \omega \cdot \frac{\partial
\cL}{\partial \Omega}+\bda {\cal B}_{(C)}\Big\}+\int_{\cR} \bda\Psi
\cdot \bda\mathcal{E}\,.
\end{equation}
The variations $\bda\Psi$ are solutions to the linearized equations
of motion $\bda \mathcal{E}=0$.

For a field theory over space and time the boundary $\bcR$ is closed
hypersurface in spacetime. Also an instant of time means a spacelike
hypersurface. Let two hypersurfaces $\Sigma_1$ and $\Sigma_2$ and
let $\cR$ be the region between them. The boundary of $\cR$ is
closing with a timelike hypersurface at spatial infinity. To bring
that a little closer let two spacelike hypersurfaces
$\Sigma^\alpha_1$ and $\Sigma^\alpha_2$ whose boundaries
$S_1^\alpha$ and $S^\alpha_2$ are connected by a (mostly) timelike
hypersurface $T^\alpha$ such that as $\alpha \to \infty$ the
boundaries $S^\alpha$ approach spheres at spatial infinity.

In order to obtain a Poincare invariant symplectic structure in
field theory we need a `no-leaking' condition: The part of the
boundary integral over $T^\alpha$ in (\ref{second-variation}) goes
to zero as $\alpha\to\infty$:
\begin{equation}\label{no-leaking-basic}
i^*_{T^\alpha}\Big(\bda\Big\{\bda \omega \cdot \frac{\partial
\cL}{\partial \Omega}+\bda {\cal B}_{(C)}\Big\}\Big) \to 0\,,
\end{equation}
as $\alpha \to \infty$. To proceed from this point one should
determine and define carefully the asymptotic symmetries at spatial
infinity, define the space of fields whose asymptotics preserve
these symmetries, and make sure that they are such that the quantity
above vanishes in the limit. This is already quite involved in
Einstein gravity and will not try to be rigorous; to obtain a result
we have to proceed somehow else. We shall assume that there are
asymptotic conditions on the fields such that the `no-leaking'
condition (\ref{no-leaking-basic}) holds and see what do we get from
that.

Condition (\ref{no-leaking-basic}) implies that locally in phase
space
\begin{equation}\label{no-leaking-exact}
i^*_{T^\alpha}\Big(\bda \omega \cdot \frac{\partial \cL}{\partial
\Omega}+\bda {\cal B}_{(C)}\Big) \to i^*_{T^\alpha} \bda B_\infty\,,
\end{equation}
in the limit $\alpha \to \infty$, for some phase space function
$B_\infty$. Define a `regularized' boundary form: ${\cal
B}_{(C)}^{\textit{reg}}:={\cal B}_{(C)}-B_\infty$. Then
\begin{equation}\label{no-leaking-exact-special}
i^*_{T^\alpha}\Big(\bda \omega \cdot \frac{\partial \cL}{\partial
\Omega}+\bda {\cal B}^\textit{reg}_{(C)}\Big) \to 0\,,
\end{equation}
as $\alpha \to \infty$. Upon replacing ${\cal B}_{(C)}$ with this
boundary form from the level of the Lagrangian (\ref{action_in_R}),
we work with a boundary form such that the `no-leaking' conditions
holds in the form (\ref{no-leaking-exact-special}).

Before proceeding let us digress briefly for some enlightening
comments. First of all let's return to equation
(\ref{first-variation}). If we restrict ourselves to variations
$\bda=\bda_{(C)}$ which respect the conditions $C$ in the given
region, by the very definition of the boundary form we have
\begin{equation}\label{C-conditions}
i^*_{\|} \left(\bda_{(C)} \omega \cdot \frac{\partial \cL}{\partial
\Omega}+\bda_{(C)} {\cal B}_{(C)}\right)=0\,,
\end{equation}
and $S$ attains an extremum for $\mathcal{E}=0$ in the interior of
the region.

Dirichlet boundary conditions amount to consider variations such
that the induced fields are held fixed. That means that ${\cal
B}_{(C)}$ is defined within a class of boundary forms which satisfy
(\ref{C-conditions}) up an arbitrary functional of the induced
fields. The idea is that one might be able to find within that class
a boundary form, denoted by ${\cal B}_{(C)}^\textit{reg}$, such that
the stronger `no-leaking' condition
(\ref{no-leaking-exact-special}).

Other boundary conditions such that a boundary form satisfying both
(\ref{C-conditions}) and (\ref{no-leaking-exact-special}) are also
possible. They are discussed in section \ref{Different boundary
conditions}.

If $\bda_{(C)}$ are replaced with general variations the r.h.s. of
(\ref{C-conditions}) does not vanish. Instead defines the
`quasi-local' energy-momentum and spin tensors
$\Theta_{\textit{quasi-local}}$:
\begin{equation}\label{quasilocal}
\bda \omega \cdot \frac{\partial \cL}{\partial \Omega}+\bda {\cal
B}_{(C)}^{\textit{reg}}=\bda \Psi \cdot
\Theta_{\textit{quasi-local}}\,,
\end{equation}
where the pullback into the boundary is understood. The terminology
comes from the fact that for Dirichlet boundary conditions
$\Theta_\textit{quasi-local}$ is indeed the quasi-local energy
tensors of Brown and York~\cite{Brown:1992br}. The modification to
the `regularized' boundary form originated from the \textit{AdS}/CFT
correspondence~\cite{Maldacena:1997re} whose framework suggests that
local boundary counter-terms should cure
divergencies~\cite{Witten:1998qj}. It was then applied successfully
to \textit{AdS} mass calculations in \cite{Balasubramanian:1999re}
following also Ref.~\cite{Henningson:1998gx}. (It was also shown
that by the same formal methods one can obtain results for de Sitter
spacetimes~\cite{Balasubramanian:2001nb} though there are minus
signs ambiguities due to nature of \textit{dS}
space~\cite{Myung:2001ab}\cite{Ghezelbash:2001vs}).


The point is that if the quasi-local tensors are finite then indeed
the r.h.s. of (\ref{quasilocal}) goes to zero at infinity, as $\bda
\Psi$ vanishes in that limit for any definition of the function
space of the fields. In other words the `no-leaking' condition
(\ref{no-leaking-exact-special}) is satisfied. In other words the
symplectic form is Poincare invariant.

The basic problem then is whether we know how to construct the form
$B^\textit{reg}_{(C)}$ such that (\ref{no-leaking-exact-special})
holds. It will be quite useful in stating certain simple
observations in this work to distinguish between the following three
cases.


Type $(\textbf{I})$. We know how to construct
$B^\textit{reg}_{(C)}$. Then the conserved charges quasi-local in
nature. Everything depends on the finiteness of the quasi-local
tensor. This guaranties that (\ref{no-leaking-exact}) can be
satisfied within a reasonable field space i.e. the asymptotics of
the field $\Psi$ will allow for variations of the parameters of the
solutions in $\bda \Psi$. Stronger conditions on asymptotics would
invalidate the whole analysis. We are clearly very heuristic here.

Type $(\textbf{II})$. We do not know how to write down a simple
enough $B^\textit{reg}_{(C)}$. This is especially the case of
asymptotically flat spacetimes.

Of course, the `we know' and `we do not know' part in the above
rather sketchy definitions depends on assumptions. For example, for
the asymptotically flat case, fairly complicated boundary terms have
been presented in works such as
\cite{Lau:1999dp,Mann:1999bt,Mann:1999pc,Kraus:1999di} which do
produce convergent results. These are related to works such as
\cite{Brown:1998bt} where another boundary term is constructed by a
`light-cone reference'; this term produces a quasi-local mass with
nice properties in the `small sphere limit'. We will not attempt to
add something new on these matters so we have nothing to offer in
those directions.

We would like, among other things, to obtain values for the
conserved charges by simple prescriptions in a uniform way in the
various dimensions and theories (of interest). The boundary forms
$B^\textit{reg}_{(C)}$ shall be the simplest polynomials of forms
constructed out of the vielbein and connections one-forms their
derivatives. This is adequate in many cases, as far as the values
for the charges are concerned. This is particularly useful in
Lovelock gravity where formulas can get a lot messier than in
Einstein gravity. The use of differential forms notation throughout
is also a choice made in that spirit.

Type $(\textbf{H})$ (`H' stands for `Hybrid'). This is a special
case of $(\textbf{I})$ such that no `corrections' are required: a
minimal boundary form consistent with the boundary conditions is
adequate for convergence.

The following definition will be completely clear after the formulas
have been derived.
\begin{definition}\label{BY-relative-definition}
The first two cases correspond to two different derivations of the
Hamiltonian generators. If a Hamiltonian is calculated as type
$(\textbf{I})$ it will be called a \textit{Brown-York} Hamiltonian.
If it is calculated as a type $(\textbf{II})$ it will be called a
\textit{relative} Hamiltonian generator. As the latter is defined
through weaker conditions a Brown-York generator can always be
thought of as a relative generator; one may have only fix a
$\bda$-integration constant appropriately. The hybrid case type
$(\textbf{H})$ is one such that a relative generator is also a
Brown-York generator.
\end{definition}
In what follows the derivations are not essentially different in
spirit from the covariant phase space methods of Wald and
collaborators~\cite{Lee:1990nz,Wald:1993nt,Iyer:1994ys,Iyer:1995kg}
though the style is closer to that of Ref.~\cite{Crnkovic:1986ex}.
The difference lies in we work throughout with first order formalism
which allows us to obtain the results in a closed form for the
Lagrangians under consideration, and in that we take the existence
of a well defined Poincare invariant symplectic form as the guiding
principle of the derivations.


\subsection{The Hamiltonian generator in type $(\textbf{I})$}
\label{section-type-I}

Suppose we know how to construct explicitly a boundary form ${\cal
B}_{(C)}$ such that (\ref{no-leaking-exact-special}) holds.
\begin{note}
From now on we drop the index `\textit{reg}' for a boundary form
satisfying (\ref{no-leaking-exact-special}).
\end{note}
We have a candidate for a symplectic form:
\begin{equation}\label{presymplectic-form}
\bds:=\bda\int_{\Sigma}\left\{ \bda \omega \cdot \frac{\partial
\cL}{\partial \Omega}+\bda {\cal B}_{(C)} \right\}\,,
\end{equation}
i.e. one that is independent of the spacelike hypersurface $\Sigma$.
Finiteness is also required but will be left as an assumption. It
will be rather clear form the very conditions imposed.

$\bds$ is a phase space 2-form which is by construction closed under
the exterior derivative: $\bda \bds=0$. $\bds$ it is not a
symplectic form because it is degenerate: it vanishes when
$\delta\Psi$ are gauge transformations. We see below that $\bds$
will naturally project into a subspace normal to the gauge
transformation directions.

Let $\xi$ be a vector field over the spacetime manifold $M$. Then
$\pounds_\xi \Psi$ can be regarded as a tangent vector over the
phase space. Let a phase space function $H_\xi$ such that
\begin{equation}\label{hamiltonian}
\boldsymbol{i}_{\pounds_{\xi}\Psi} \bds=-\bda H_\xi\,.
\end{equation}
($\boldsymbol{i}_{\boldsymbol{\Xi}}$ is the inner product on forms
in phase space with the phase space vector $\boldsymbol{\Xi}$.)
$H_\xi$ is the Hamiltonian generator in phase space, or simply the
Hamiltonian, of the transformations generated by $\xi$ in spacetime.
Relations (\ref{hamiltonian}) are Hamilton's equations.

Here we got an action and a symplectic form $\boldsymbol{\sigma}$
but not a Hamiltonian. The transformations generated by $\xi$, the
diffeomorphisms, are a symmetry of the action and therefore of
$\boldsymbol{\sigma}$. The Hamiltonian can be constructed through
the existence of the symmetry as a momentum mapping~\cite{marsden}
essentially using a Noether type method.

If $\bds$ is invariant along a vector $\boldsymbol\Xi$ in phase
space this formally means that
$\boldsymbol\pounds_{\boldsymbol\Xi}\bds=0$. By the closure of
$\bds$ this means that $\bda
\left(\boldsymbol{i}_{\boldsymbol\Xi}\bds\right)=0$. We used
Cartan's relation for the Lie derivative in phase space:
$\boldsymbol\pounds_{\boldsymbol\Xi}=\bda\boldsymbol{i}_{\boldsymbol\Xi}+\boldsymbol{i}_{\boldsymbol\Xi}\bda$.
Thus the phase space one-form $\boldsymbol{i}_{\boldsymbol\Xi}\bds$
must be exact: $\boldsymbol{i}_{\boldsymbol\Xi}\bds=-\bda H$, for
some phase space function $H$. That this holds globally is an
assumption. Now if the `Hamiltonian' $H$ is given, and the $\bds$ is
non-degenerate, then the last relation defines a Hamiltonian vector
field $\boldsymbol{\Xi}_H$. Hamilton's equations hold along its
integral curves $\pounds_\xi \Psi=\boldsymbol{\Xi}_H$.

Now $\bds$ being diffeomorphism invariant means that it does change
along the directions on the tangent of the phase space that
correspond to diffeomorphism transformations of the fields. That is
$\boldsymbol\pounds_{\pounds_\xi \Psi}\bds=0$. Through a reasoning
similar to the above, we arrive at (\ref{hamiltonian}) for some
phase space function $H_\xi$ which we must construct, the
Hamiltonian generators of transformations over $M$ generated by
$\xi$.

A debt left from earlier is that $\bds$ does project on the space of
solutions modulo gauge transformations, by having no components in
those directions. This statement translates to:
$\boldsymbol{i}_{\pounds_{\xi}\Psi} \bds=0$. This is the case when
$\xi$ has a compact support in $M$, or better corresponds to proper
gauge transformations in the language
of~\cite{Regge:1974zd}\cite{Benguria:1976in}. Then $H_\xi$ like all
gauge symmetry generators vanish identically on-shell, as generators
of local symmetries should do, and $\bds$ projects. When $\xi$ are
non-trivial asymptotically, are improper diffeomorphisms, the
generators become non-trivial tensorial quantities of the asymptotic
symmetry groups. They should of course be related to the asymptotic
values of the fields.

Regge-Teitelboim~\cite{Regge:1974zd} produce the generators starting
with the Hamiltonian as the usual Legendre transform of the
Lagrangian, and postulating additional suitable boundary integrals
which make all the required functional derivatives well defined.
There is quite a bit of analogy between that construction and
covariant phase space methods which work with action functional,
only in the latter the canonical variables are not explicitly
picked. The final on-shell must agree wherever the methods make
sense though the results may differ by phase space constants i.e.
parameters that depend on the theory (couplings) and not the
solution. This happens explicitly in the case of locally flat
spacetime in three dimensions, as well shall see.

Finally, without dwelling further into the symplectic form of the
Hamiltonian theory we will find an $H_\xi$ starting from
(\ref{hamiltonian}) and (\ref{presymplectic-form}). We have
\begin{equation}\label{}
\bds=\int_\Sigma \bda \omega \cdot \bda \frac{\partial {\cal
L}}{\partial\Omega}\,.
\end{equation}
We have
\begin{equation}\label{}
\boldsymbol{i}_{\pounds_{\xi}\Psi} \bds=\int_\Sigma \pounds_\xi
\omega \cdot \bda \frac{\partial {\cal L}}{\partial\Omega}+
\bda\omega \cdot \pounds_\xi \frac{\partial {\cal
L}}{\partial\Omega}\,.
\end{equation}
By $\bda\pounds_\xi=\pounds_\xi \bda$ the integrant can also be
written as
\begin{equation}\label{i-sigma-integrant}
-\bda \left\{\pounds_\xi \omega \cdot \frac{\partial {\cal
L}}{\partial\Omega} \right\}+\pounds_\xi \left\{\bda \omega \cdot
\frac{\partial {\cal L}}{\partial\Omega} \right\}\,.
\end{equation}
Using $\bda\Omega=-D\bda\omega$ and $\pounds_\xi \omega=D(i_\xi
\omega)+i_\xi \Omega$ one can easily show that the first bracket
reads
\begin{equation}\label{identity}
d \left(i_\xi \omega \cdot \frac{\partial {\cal L}}{\partial\Omega}
\right)+i_\xi {\cal L}-i_\xi \Psi \cdot \mathcal{E}\,.
\end{equation}
This is a quite useful fact. Then it is easy to show that the whole
of (\ref{i-sigma-integrant}) reads
\begin{align}\label{}
& -\bda d \left(i_\xi \omega \cdot \frac{\partial {\cal
L}}{\partial\Omega} \right)+d i_\xi \left(\bda \omega \cdot
\frac{\partial {\cal L}}{\partial\Omega} \right)\\
& +\bda \Psi \cdot i_\xi \mathcal{E}+i_\xi \Psi \cdot \bda
\mathcal{E}\,. \nonumber
\end{align}
$\bda i_\xi=-i_\xi \bda$ is used often. We have kept the terms
involving $\mathcal{E}$ and $\bda \mathcal{E}$ in this formula for
completeness. So we have that $\boldsymbol{i}_{\pounds_{\xi}\Psi}
\bds$ equals
\begin{equation}\label{walding}
-\bda \int_{S^\alpha}i_\xi \omega \cdot \frac{\partial {\cal
L}}{\partial\Omega}+\int_{S^\alpha} i_\xi \left(\bda \omega \cdot
\frac{\partial {\cal L}}{\partial\Omega} \right)\,,
\end{equation}
where the limit $\alpha\to\infty$ is understood.

Now $S^\alpha$ is a boundary of both $\Sigma^\alpha$ and $T^\alpha$.
We may orient the manifold $T^\alpha$ to be such that $\xi$ is
tangent to it at the boundary $S^\alpha$. Then the `no-leaking'
condition (\ref{no-leaking-exact-special}), extended to the boundary
of $T^\alpha$, tells us that
\begin{equation}\label{}
i^*_{S^\alpha} i_\xi \left(\bda \omega \cdot \frac{\partial {\cal
L}}{\partial\Omega} \right) \to -i^*_{S^\alpha} i_\xi \bda {\cal
B}_{(C)}\,,
\end{equation}
as $\alpha\to \infty$. Thus we have
\begin{equation}\label{hamilton-eq-explicit}
\boldsymbol{i}_{\pounds_{\xi}\Psi}\bds
=-\bda\left\{\int_{S^\alpha}i_\xi
\omega \cdot \frac{\partial {\cal L}}{\partial\Omega}+i_\xi {\cal
B}_{(C)} \right\}\,.
\end{equation}
By (\ref{hamiltonian}) we have
\begin{equation}\label{the-hamiltonian}
H_\xi=\int_{S^{\infty}} \left\{ i_\xi \omega \cdot \frac{\partial
{\cal L}}{\partial\Omega}+i_\xi {\cal B}_{(C)} \right\}+H^0_\xi\,.
\end{equation}
(\ref{hamiltonian}) defines the Hamiltonian up to closed zero-form
that is a constant $H^0_\xi$.

\begin{note}
It is important to remember that this formula was obtained by
treating $S^\alpha$ as boundary of $T^\alpha$. Therefore ${\cal
B}_{(C)}$ is constructed as a boundary form for the time-like
component of the boundary.
\end{note}

To recapitulate, as long as (\ref{extremized}) and
(\ref{no-leaking-exact-special}) make sense, they are all we need to
construct explicitly a Hamiltonian for $\xi$ in a fully covariant,
first order formulation. The Hamiltonian $H_\xi$ stems from the
existence of a finite quasi-local tensor, thus that notion is
underlying the derivation in an essential way but the tensor itself
is not explicitly used.

\begin{note}
We have taken for granted that the only boundary the spacelike
hypersurfaces $\Sigma$ have is a surface at large radial distances.
In case this is not true an analogous condition
(\ref{no-leaking-exact-special}) and corresponding additional terms
in $H_\xi$ must be included, a procedure not really well-defined
near actual singularities. This is a crucial detail. In what follows
we would like to have the spatial infinity as the only boundary of
$\Sigma$. For that to be so, we shall assume from now on that
singular energy distributions implied by the metrics we write down
are replaced by smooth ones such that no timelike singularities
arise.
\end{note}

The value of the `angular' integral over $S^\infty$ in
(\ref{the-hamiltonian}) cannot be shifted by addition of local term
on the boundary action terms. The phase space constant $H^0$ cannot
be shifted. Its presence makes the generator potentially consistent
with other constructions of the generators but it is a quantity
without much meaning when we discuss quasi-local generators. Thus we
shall give a name to the `angular' integral. We will refer to it as
the \textit{Brown-York} generator:
\begin{equation}\label{BY-definition}
H=H_{BY}+H^0\,.
\end{equation}

\subsection{Quasi-local generators and conservation}
\label{conservation}

We mentioned already in the introduction that a major problem in
gravity is localizing energy. All our formal choices might appear
natural from certain points of view, but they are essentially
motivated by proposals addressing the localization problem.
Equivalence principle or general covariance say that any energy
density must vanish.

Going to the boundary was the brilliant move of Brown and
York~\cite{Brown:1992br}. It fits nicely in `t Hooft's `holography'
idea: gravitational degrees of freedom are fundamentally boundary
degrees of freedom~\cite{'tHooft:1993gx}. There is a host of
quasi-local definitions, see e.g. Ref.~\cite{Hayward:1993ph} and the
works discussed there. Misner-Sharp mass~\cite{Misner:1964je} is
another definition of mass of quasi-local type, applying in
spherical symmetry, with nice properties~\cite{Hayward:1994bu}; for
a Lovelock gravity it was defined recently in \cite{Maeda:2007uu}.

In our derivation we encountered the quantities
\begin{equation}\label{the-hamiltonian-quasilocal}
H_{\xi}(\alpha)=\int_{S^{\alpha}} \left\{ i_\xi \omega \cdot
\frac{\partial {\cal L}}{\partial\Omega}+i_\xi {\cal B}_{(C)}
\right\}+H_\xi^0\,.
\end{equation}
What is important about them is that they converge to the canonical
generator $H_\xi$ for $\alpha \to \infty$, where $r_\alpha$ is the
radius of the hypersurface $T^\alpha$. One may regard
$H_\xi(\alpha)$ as the value of the associated charge contained in
the spatial sections bounded by $T^\alpha$.

Though this is not precise, as we will discuss in section
\ref{interpolations}, there is a reason for thinking something like
that. $H_\xi(\alpha)$ is essentially the quasi-local mass of Brown
and York, in first order formalism; their mass is defined as a
canonical generator in the sense of Hamilton-Jacobi theory
\cite{Brown:1992br}\cite{Brown:2000dz}. In order to understand the
claim, let $T^\alpha$ be a timelike hypersurface and $\zeta$ be a
vector field tangential to it. $\zeta$ is not a Killing vector. By
manipulations learned in the previous section it is straightforward
to show that
\begin{align}
&\int_{T^\alpha}d\Big\{i_\zeta\omega \cdot \frac{{\partial\cal
L}}{\partial \Omega}+i_\zeta {\cal B}_{(C)} \Big\}=\\
&\int_{T^\alpha}\Big\{\pounds_\zeta\omega \cdot \frac{{\partial\cal
L}}{\partial \Omega}+\pounds_\zeta {\cal B}_{(C)}-i_\zeta ({\cal
L}+d{\cal B}_{(C)})+i_\zeta\Psi \cdot {\cal E}\Big\}\,. \nonumber
\end{align}
The last term vanishes on-shell, ${\cal E}=0$. The middle term
vanishes as $\zeta$ is tangent on $T^\alpha$. Now, note first that
the remaining term is nothing but the boundary term on $T^\alpha$ of
the action functional for Lie derivative variations of the fields,
see (\ref{first-variation}). Also we see from (\ref{quasilocal})
that this reads $\pounds_\zeta \Psi \cdot
\Theta_\textit{quasi-local}$. From precisely this term the
quasi-local densities are read off from in the Hamilton-Jacobi
analysis of Refs.~\cite{Brown:1992br}\cite{Brown:2000dz}. Now, the
l.h.s. of the equation is nothing but the difference of
$H_\zeta(\alpha)$ evaluated in the future and past $S^\alpha$. In
other words, we obtain explicitly that the integrand involving the
Brown-York quasi-local densities is a total derivative producing
$H_\zeta(\alpha)$ on the boundary sphere. Thus, as mentioned in the
introduction, $H_\zeta(\alpha)$ is essentially the canonical
generator in the sense of Hamilton-Jacobi theory.

Conservation is easily obtained from the previous formula if we let
$\zeta$ be a Killing vector: then the future and past values of
$H_\zeta(\alpha)$ must agree. The time-derivative of
(\ref{the-hamiltonian-quasilocal}) can also be calculated directly.
Let a $\chi$ be a time-like vector field tangential on $T^\alpha$.
By $\partial S^\alpha=0$ the previous formula tells us that the rate
of change $\dot H_\xi(\alpha)$ of the quasi-local charge
$H_\xi(\alpha)$ under the flow of $\chi$ reads:
\begin{align}
\dot H_\xi(\alpha)=\int_{S^\alpha} & \Big\{i_\chi \big\{ \pounds_\xi
\omega \cdot
\frac{\partial {\cal L}}{\partial \Omega}+\pounds_\xi {\cal B}_{(C)}\big\} \nonumber\\
&-i_\chi i_\xi ({\cal L}+d{\cal B}_{(C)})+i_\chi(i_\xi \Psi \cdot
{\cal E})\Big\}\,.
\end{align}
For any Killing vector $\xi$ we have that $\pounds_\xi \Psi=0$ thus
the first term should vanish. (If $\xi$ is an asymptotic Killing
vector then conservation holds in the limit $\alpha \to \infty$.) If
$\xi$ is a time-like Killing vector the remaining term vanishes for
$\chi=\xi$ by the identity $i_X i_X=0$. (In general, the inner
product operator satisfies: $i_X i_Y=-i_Y i_X$ for any vector fields
$X$ and $Y$.) This a remnant of the `commutativity' of Hamiltonian
with itself. Mass and angular momentum as given by
(\ref{the-hamiltonian-quasilocal}) are indeed conserved on-shell.

\subsection{Different boundary conditions}
\label{Different boundary conditions}

A question which arises is: To what extent the Hamiltonian depends
on the conditions $C$? To answer that one should look at the
associated no-leaking condition, equation
(\ref{no-leaking-exact-special}), which we may call it $F(C)$. The
phase space is defined such that this condition holds. Let us denote
by $C_1$ and $C_2$ two different boundary conditions. one works in
two phase spaces where the conditions $F(C_1)$ and $F(C_2)$ hold.
Two things may happen. One, the phase spaces have a measure zero
intersection. Then we may obtain two different Hamiltonians. Second,
the intersection of phase spaces is nice subset of each one. Then we
may restrict our variations in there. Subtracting the `no-leaking'
conditions (\ref{no-leaking-exact-special}) we have that
$i^*_{T^\alpha} (\bda {\cal B}_{(C_1)}-\bda {\cal B}_{(C_2)})$ goes
to zero for $\alpha\to\infty$. The previous derivation shows that
the two Hamiltonians (\ref{the-hamiltonian}) associated with $C_1$
and $C_2$ may differ only by a phase space constant. For example, a
set of counter-terms applying to asymptotically \textit{AdS}
spacetimes were constructed and applied in
Refs.~\cite{Mora:2004kb,Mora:2004rx,Kofinas:2006hr,Olea:2006vd,Miskovic:2007mg,Kofinas:2007ns}
as boundary terms respecting the conformal symmetry of \textit{AdS}
boundary. Though those counter-terms differ a lot from the simple
boundary forms we are using the results are expected to agree and
they indeed do as we shall see in examples. Additionally the same
boundary forms apply straightforwardly with little formal
modification to asymptotically de Sitter spacetimes, along the lines
of Ref.~\cite{Balasubramanian:2001nb}. From the point of view of
Refs.~\cite{Mora:2004kb,Mora:2004rx,Kofinas:2006hr,Olea:2006vd,Miskovic:2007mg,Kofinas:2007ns}
it is not clear why this is so; from the point of view of the
Dirichlet counter-terms it is not that surprising.

\subsection{The Hamiltonian generator in type $(\textbf{II})$}
\label{The Hamiltonian generator in type II}

When spacetime is asymptotically flat, there is no length scale to
facilitate regularization. It is then characteristically difficult
to construct a boundary form ${\cal B}_{(C)}^\textit{reg}$ that
would lead to finite charges. On the other hand, in this case the
ADM construction of charges~\cite{Arnowitt:1959ah} it is a
consistent method for calculating the energy and the angular
momentum for asymptotically flat space. An extension of this is the
Abbott-Deser mass~\cite{Abbott:1981ff} which calculates total energy
for asymptotically de Sitter spacetimes.

From the point of view of first order formulation these are all
subtraction methods. Energy and any charge is measured
\emph{relatively} to background by studying and formulating things
in terms of the asymptotic deviations from the background.

In such cases the form ${\cal B}^\textit{reg}$ should exist but one
cannot find a local expression of it, or doesn't care to find one.
It is not an accident that this applies to asymptotically flat
spaces: the `holographic renormalization' does not work
straightforwardly in asymptotically flat spacetimes and the
`corrections' need to be non-local~\cite{Skenderis:2002wp}. In
specific case on though may find local but not polynomial
counterterms which suffice to remove divergencies but they do not
apply to all asymptotically flat
spaces~\cite{Lau:1999dp,Mann:1999bt,Mann:1999pc,Kraus:1999di}. For
the general case, non-locality suggests that the boundary form can
be constructed by involving \emph{another} solution of the equations
of motion i.e. another point in phase space.

Let $\Psi$ and $\Psi_\infty$ be two solutions of the the equations
of motion. The indication `$\infty$' is simply a name to distinguish
the second solution, its meaning will be explained below. Then
subtracting conditions (\ref{no-leaking-exact}) applied on each
solution we have:
\begin{equation}\label{exact-difference}
\left( \bda \omega \cdot \frac{\partial \cL}{\partial \Omega}+\bda
{\cal B}_{(C)} \right) -\left( \bda \omega \cdot \frac{\partial
\cL}{\partial \Omega}+\bda {\cal B}_{(C)} \right)_\infty \to \bda
B\,,
\end{equation}
pulled back into $T^\alpha$, $\alpha \to \infty$.

$B$ is a phase space function which is antisymmetric in fields
$\Psi$ and $\Psi_\infty$. $B$ must be annihilated by $\delta_{(C)}$,
that is, $\delta B=\delta B(i^*\Psi,i^*\Psi_\infty)$. This is an
important fact: Choose a field $\Psi_\alpha$ to induce the same
field with $\Psi$ into $T^\alpha$:
\begin{equation}\label{matching}
i_{T^\alpha}^*\Psi=i_{T^\alpha}^*\Psi_\alpha\,,
\end{equation}
such that there is a well defined limit $\Psi_\alpha \to
\Psi_\infty$ as $\alpha \to \infty$. Then by antisymmetry $\delta
B=0$ i.e.
\begin{equation}\label{no-difference}
\bda \omega \cdot \frac{\partial \cL}{\partial \Omega} \to -\bda
{\cal B}_{(C)} + \left( \bda \omega \cdot \frac{\partial
\cL}{\partial \Omega}+\bda {\cal B}_{(C)} \right)_\infty
\end{equation}
pulled back into $T^\alpha$, $\alpha \to \infty$. Condition
(\ref{matching}) is not necessary but allows us solubility; without
it we are simply back to the general (\ref{exact-difference}).

Let find now an explicit form for the $\infty$-part of the r.h.s. of
(\ref{no-difference}) which we know is $\bda$-exact. Let
$\Psi_\infty$ be any solution that can be continuously related to
$\Psi$ by the parameters. We have
$(\boldsymbol{i}_{\pounds_{\xi}\Psi} \bds)_\infty=-\bda
H_{\infty,\xi}$. Then relation (\ref{walding}) applied to
$\Psi_\infty$ allows to write the r.h.s. of (\ref{no-difference}) as
$\bda$-exact, at least under the integral sign in the process of the
derivation of $H_\xi$. This is all one needs. Applying
(\ref{no-difference}) in the derivation at the step of
(\ref{walding}) one finds easily
\begin{align}\label{poor-mans-hamiltonian}
H_\xi=\int_{S^{\infty}}  &\left\{i_\xi \omega \cdot \frac{\partial
{\cal L}}{\partial\Omega}+i_\xi {\cal B}_{(C)}\right. \nonumber
\\ &
\quad - \left. \left(i_\xi \omega \cdot \frac{\partial {\cal
L}}{\partial\Omega}+i_\xi {\cal B}_{(C)} \right)_{\!\infty}
\right\}+H_{\infty,\xi}\,.
\end{align}
The case $\Psi=\Psi_{\infty}$ tells us that the $\bda$-integration
constant is zero. This is our Hamiltonian for the cases
$(\textbf{II})$ e.g. when spacetimes are asymptotically flat. It is
derived under the condition (\ref{no-leaking-exact}) in field space
and condition (\ref{matching}) for the second field $\Psi_\infty$.

Unlike the constant $H^0_\xi$ in (\ref{the-hamiltonian}),
$H_{\infty,\xi}$ is not a constant function in phase space. It is
the value at of the Hamiltonian at $\Psi_\infty$. Now $\bda
H_{\infty,\xi}$ does not vanish unless $\Psi_{\infty}$ is
equilibrium point of the phase space. This is true if $\Psi_\infty$
is an actual `ground state' of the system. Of course in any case the
$H_\xi$ satisfies (\ref{hamiltonian}) by construction.

Condition (\ref{matching}) is crucial for applications of
(\ref{poor-mans-hamiltonian}). In general it cannot be satisfied if
the fields are thought of as components w.r.t. the same coordinates.
The induced fields will agree on $T^\alpha$ as geometric objects up
to necessary diffeomorphisms. There is nothing unnatural in that
when using manifestly invariant quantities. A condition of the form
(\ref{matching}) was introduced in \cite{Hawking:1995fd} as a means
to guaranty finite-ness of a `physical' action
$S(\Psi)-S(\Psi_\infty)$, dragging also the conserved quantities to
finite-ness. Here it was introduced as a convenient step integrating
a Poincare invariant symplectic form in phase space.

We can now give a specific definition of the relative (on-shell
values of the) generators, first mentioned in definition
\ref{BY-relative-definition}.
\begin{definition}
\label{relative-mass-definition} Every case can be regarded as type
$(\textbf{II})$. So may bypass the arbitrariness of $H_\infty$ or
use the arbitrariness of $H^0$ in type $(\textbf{I})$ to define a
relative generator $H_{\textit{rel}}$ such that: $H_\textit{rel}=0$
if a certain condition holds. The condition usually expresses the
fact that we measure the charges such that their vacuum values are
set to zero.
\end{definition}


\section{Spherical and `topological' metrics}
\label{Spherically symmetric metrics}

We start by studying some characteristic examples in three, four and
five spacetime dimensions in Einstein gravity. Some interesting
things arise when the angular manifolds are not spheres. Einstein
gravity will be described by the Lagrangian
\begin{equation}\label{einstein-lagrangian}
{\cal L}=\frac{1}{2}c_1 \epsilon(\Omega \underbrace{E \cdots
E}_{D-2})=\frac{1}{2}c_1 \epsilon(\Omega E^{D-2})\,,
\end{equation}
where $(D-2)! c_1=(8\pi G)^{-1}$, where $G$ we define as Newton's
constant in the given dimensions, following the most standard
conventions.

The standard form of the metrics we consider is
\begin{equation}\label{general-element}
ds^2=-g^2 dt^2+\frac{dr^2}{g^2}+r^2 d\Omega^2\,,
\end{equation}
and $d\Omega^2$ is the metric of a compact i.e. closed, constant
curvature co-dimension 2 manifold. Let a frame $\tilde E^i$ over
this manifold i.e. $d\Omega^2=\delta_{ik} \tilde E^i \otimes\tilde
E^k$ and $\tilde\omega^i_{~j}$ a connection. Its curvature will be
$\tilde\Omega^{ij}=k\,\tilde E^i\tilde E^j$, $k=\pm 1, 0$. Of course
$k=1$ refers to the spherically symmetric case. But interesting
facts are related to comparison of the that case to the flat ($k=0$)
manifold. The black holes associated with $k$ different than one
have been termed as `topological' (see below).

Let the frame
\begin{equation}
E^0=g\,dt\,, \quad E^1=g^{-1}dr\,, \quad E^i=r\, \tilde E^i\,.
\end{equation}
Then of course $ds^2=\eta_{AB} E^A\otimes E^B$. As long as torsion
is zero, and this always the case for Einstein gravity in vacuum, we
have that the non-zero components of the connection are
\begin{equation}\label{}
\omega^0_{~1}=gg'\,dt\,, \quad \omega^i_{~1}=g\, \tilde E^i\,, \quad
\omega^i_j=\tilde \omega^i_{~j}\,.
\end{equation}

The first one we would like to analyze is the most standard of all:
four dimensional Schwarzschild metric with zero cosmological
constant. Then
\begin{equation}\label{schwarzschild}
g^2=1-\frac{2Gm}{r}\,.
\end{equation}
We would like to analyze how $m$ arises as the total energy of the
system.

Energy, or mass, will be calculated as the on-shell value of the
Hamiltonian for the Killing vector $\xi=\partial/\partial t$. The
problem, due to asymptotic flatness, is case $(\textbf{II})$ as one
may check for oneself. We use (\ref{poor-mans-hamiltonian}).

For Einstein gravity and Dirichlet boundary conditions the boundary
form is ${\cal B}_{(C)}=-\frac{1}{2}c_1\, \epsilon(\theta EE)$, as
discussed in the appendix. The form $\theta=\omega-\omega_{\|}$ is
constructed from the induced connection $\omega_{\|}=i^*\omega^{ab}$
of a constant $r=r_\alpha$ hypersurface $T^\alpha$, where $a,b=0,i$.
[The induced and intrinsic fields equalities, for the connection as
well as $E_\|=i^* E$, are consistent for zero intrinsic torsion.]

Thus we have $\theta^{01}=gg' dt$ and $\theta^{i1}=g \tilde E^i$.
The first part of the integral in (\ref{poor-mans-hamiltonian})
reads
\begin{equation}\label{schwarzschild-hamiltonian}
\int_{S^2} c_1 \cdot \left(\frac{1}{2} \epsilon(i_\xi \omega
EE)-\frac{1}{2} i_\xi \epsilon(\theta EE)\right)=-\frac{1}{G}g^2
r\,,
\end{equation}
evaluated at $r=r_\alpha \to \infty$ as $\alpha \to \infty$. This
term of (\ref{poor-mans-hamiltonian}) alone diverges. Before
proceeding we digress for a couple of comments.

One should note that there are no derivatives in the result. The
divergence is due to this fact. This means that the `no-leaking'
condition is not respected thus (\ref{schwarzschild-hamiltonian}) is
not a final result. Nonetheless, we may note that in fact all our
final formulas will ot contain derivatives of the metric function.

Ref.~\cite{Ashtekar:2008jw} seems to have a general setting similar
to ours. Various ingredients e.g. first order formalism, boundary
forms, covariant phase space are there. The results though disagree,
the Hamiltonian is convergent by construction. The difference lies
in the definition of the phase space. \cite{Ashtekar:2008jw} imposes
conditions on the fields associated to asymptotic flatness. Without
them certain ambiguities of the asymptotic Poincare group i.e of the
energy-momentum definitions arise in four dimensions. Doing so, one
again implicitly works with Minkowski as reference spacetime from
the very beginning, trying to make sense of the formulas under that
condition. Moreover, the asymptotic conditions on $\omega$ are
imposed using torsion-free-ness. This brings the analysis closer to
the second order formalism. In all, the results of
Ref.~\cite{Ashtekar:2008jw} are expected to be very different from
ours. The ambiguities themselves are a matter of geometry first of
all, and any set of requirements on the fields which remove them
could be applied
when a definition of energy has been adopted. We will not go into
these issues here.

We introduce fields $\Psi_\alpha$ such that the condition
(\ref{matching}) holds: Let
\begin{equation}\label{general-element}
ds^2_\alpha=-g_\alpha^2
dt^2_\alpha+\frac{dr_\alpha'^2}{g_\alpha^2}+r_\alpha'^2 d\Omega^2\,,
\end{equation}
where $g_\alpha=g_\alpha(r_\alpha')$ is another solution, say for
the parameter $m_\alpha$. The change of coordinates is given by
$r=r_\alpha$, $r_\alpha'=r_\alpha$ and $g(r_\alpha)
t=g_\alpha(r_\alpha) t_\alpha$ for some fixed radius $r_\alpha$.
Thus (\ref{matching}) hold.

What in particular (\ref{matching}) means is that
\begin{equation}
\xi=\frac{\partial}{\partial
t}=\frac{g}{g_\alpha}\,\frac{\partial}{\partial t_\alpha}\,.
\end{equation}
Thus the integral in the Hamiltonian (\ref{poor-mans-hamiltonian})
equals
\begin{equation}\label{HH}
-\frac{1}{G}(g^2 r-g g_\alpha r)\,,
\end{equation}
everything evaluated at $r_\alpha$ which goes to infinity. This is
convergent and we have $H_\xi=m-m_\infty+H_{\infty,\xi}$. In
particular we may use Minkowski spacetime fields $m_\infty=0$ as the
$\Psi_\infty$ phase space point. As mentioned in the previous
section, for the asymptotically non trivial Killing $\xi$ the
corresponding canonical generators are tensorial quantities of the
asymptotic symmetries. In particular as long as these symmetries
indeed are Lorentz transformations then $H_{\infty,\xi}=0$ that is
$H_\xi=m$ indeed.
\\

Analyzing how the result arises one observes something interesting.
The finite part of (\ref{schwarzschild-hamiltonian}) is $2m$. The
divergent part is due to the angular manifold (2-sphere) constant
curvature: $\tilde\Omega^{ij}=1\cdot \tilde E^i \tilde E^j$.
Consider now the more general case given by (\ref{general-element})
i.e. $k=\pm 1, 0$. Then the field equations are equivalent to $(r
g^2)'=k$, that is
\begin{equation}\label{schwarzschild-k}
g^2=k-\frac{r_0}{r}\,,
\end{equation}
where $r_0$ is an integration constant. These solutions are the zero
cosmological constant analogues of the \textit{AdS} topological
black
holes~\cite{Mann:1996gj}\cite{Brill:1997mf}\cite{Vanzo:1997gw}.

When $k=0$ the two-manifold is flat. It can be a torus or a Klein
bottle. Its volume $\textit{vol}_0$ is arbitrary. In that case
(\ref{schwarzschild-hamiltonian}) \emph{converges}. That is the
problem is a type $(\textbf{H})$ case. The Brown-York mass reads:
\begin{equation}\label{k=0}
k=0\: : \quad m_{BY}=\frac{\textit{vol}_0}{4\pi}\frac{r_0}{G}\,.
\end{equation}
Let's re-calculate the mass as a relative mass applying the type
$\textbf{II})$ Hamiltonian (\ref{poor-mans-hamiltonian}) which here
reads (\ref{HH}). Let $g_\alpha$ be such that $r_0(\alpha) \to 0$
for $\alpha \to \infty$. Clearly the result is again (\ref{k=0}).
Thus we indeed deal with a relative mass which can also be regarded
as a Brown-York mass, in accordance with the definition
\ref{BY-relative-definition} of the type $(\textbf{H})$ case.

This phenomenon persists even in the quite more complicated case of
Einstein-Gauss-Bonnet gravity as we will see. This is due to the
fact that $g_\alpha\to 0$ for the specific solution. There is a
finite discontinuity going from the \textit{AdS} metrics of length
scale $l$ to their $l=\infty$ analogues when they are $(\textbf{H})$
metrics, as we shall see.
\\

Let us turn to study anti de Sitter vacuum metrics in three, four
and five dimensions, in the presence also of mass. An anti de Sitter
vacuum means that the equations of motion read
$\epsilon(\{\Omega+l^{-2}EE\}E^{D-3}\delta E)=0$, that is the bulk
Lagrangian is
\begin{equation}\label{}
{\cal L}=\frac{1}{2}c_1\epsilon(\Omega E^{D-2})+\frac{1}{2}c_1
\frac{D-2}{l^2 D}\epsilon(E^D)\,.
\end{equation}
Again $(D-2)!c_1=(8\pi G)^{-1}$.

The three dimensional pure anti de Sitter spacetime is of the form
(\ref{general-element}), explicitly
\begin{equation}\label{global 3-ads}
ds^2=-\Big(1+\frac{r^2}{l^2}\Big)dt^2+\Big(1+\frac{r^2}{l^2}\Big)^{-1}dr^2+r^2d\phi^2\,.
\end{equation}
The vielbein is $E^0=g\,dt$, $E^1=g^{-1}dr$, $E^2=rd\phi$, and the
non-zero components of the connection read $\omega^0_{\ 1}=gg'dt$,
$\omega^2_{\ 1}=g\, d\phi$.

We first calculate
\begin{equation}\label{3-ads-mass-uncorrected}
\int_{S^1} c_1 \cdot \left(\frac{1}{2} \epsilon(i_\xi \omega
E)-\frac{1}{2} i_\xi \epsilon(\theta E)\right)=-2\pi c_1\, g^2\,,
\end{equation}
for $\xi=\partial/\partial t$. This expression diverges
\begin{equation}\label{}
-2\pi c_1 \frac{r^2}{l^2}-2\pi c_1\,.
\end{equation}
The problem is type $(\textbf{I})$ case. The previous result was
calculated with the minimal Dirichlet boundary form, discussed in
the appendix. Let a boundary form as
\begin{equation}\label{3 counterterm}
{\cal B}_{(C)}=-\frac{1}{2}c_1 \epsilon(\theta E)+\frac{1}{2}c_1 b_0
\epsilon(E_\| E_\|)_1\,.
\end{equation}
\begin{note}
$\epsilon(\cdots)_1$ shall mean from now on contraction with the
volume form $\epsilon_{a \dots b1}$ on the intrinsic Lorentz bundle;
it is identified with the induced `bulk' volume form, where $E^1$ in
practice is always the radial co-vector. A theoretical problem is
that $\epsilon_{a \dots b1}$ is not invariant under variations,
potentially contributing terms inexistent in our derivations. This
is not so. Let $\zeta$ a unit vector, $\eta_{AB}\zeta^A \zeta^B=1$,
normal on the boundary. That is $\epsilon_{a \dots
b1}=\epsilon_{a\dots bA}\zeta^A$. $\bda \zeta^A$ is not zero but the
unit length of $\zeta$ implies that $\bda \zeta^A$ is normal to
$\zeta^A$ i.e. tangential. As a result it drops out the variations
of the intrinsic boundary forms as it contributes one tangential
index too many.
\end{note}
The result (\ref{3-ads-mass-uncorrected}) is `corrected' by $-2\pi
c_1 b_0\, g r$ which asymptotically reads
\begin{equation}\label{}
-2\pi c_1 b_0 \frac{\ r^2}{l}-\pi c_1 b_0\, l
\end{equation}
Canceling the divergence fixes $b_0=-l^{-1}$. So finally we have
\begin{equation}\label{final 3-ads mass}
m_{BY}=-c_1\, \pi=-\frac{1}{8G}\,.
\end{equation}
Expression (\ref{final 3-ads mass}) is a well known result.

For the Poincare patch of the \textit{AdS}$_3$
\begin{equation}\label{}
-\frac{\ r^2}{l^2}dt^2+\frac{\ l^2}{r^2}dr^2+r^2d\phi^2
\end{equation}
one similarly finds the correct answer $m_{BY}=0$. Things work
nicely. The three dimensional case will be discussed and commented
on in detail in the next section.

A four dimensional anti de Sitter Schwarzschild solution is given
the $k=1$ case of (\ref{general-element}) with
\begin{equation}\label{}
g^2=\frac{r^2}{l^2}+k-\frac{r_0}{r}\,.
\end{equation}
where $l$ is the \textit{AdS} radius and $r_0$ an integration
constant. We will consider the general $k$. This is a case
(\textbf{I}) and we apply (\ref{the-hamiltonian}) for a
(non-minimal) boundary form:
\begin{equation}\label{}
{\cal B}_{(C)}=\frac{1}{2}c_1 \left\{-\epsilon(\theta EE)+b_0
\epsilon(E_\|^3)_1+b_1 \epsilon(\Omega_\| E_\|)_1 \right\}\,.
\end{equation}
The constants are to be determined by finiteness. The Hamiltonian is
given by the result (\ref{schwarzschild-hamiltonian}) plus
$b$-`corrections':
\begin{equation}\label{}
-16\pi c_1\, g^2 r+ 12 \pi c_1 b_0\, gr^2+4 \pi c_1 b_1\, kg\,,
\end{equation}
times a factor $\textit{vol}_k/(4\pi)$ where $\textit{vol}_k$ is the
volume of the 2-manifold. The expression above is evaluated at a
constant $r=r_\alpha$ which goes to infinity. The divergencies are
canceled for $b_0=4/(3l)$ and $b_1=2l$. $b_1$ is useless and not
defined for $k=0$. The Brown-York Hamiltonian turns out to be
\begin{equation}\label{}
m_{BY}=\frac{\textit{vol}_k}{4\pi}\frac{r_0}{2G}\,.
\end{equation}
This holds for any $k$ and regularization is required in all cases.

For the spherical case $k=1$ this is the well known Schwarzschild
mass we found in the asymptotically flat case, formally $l \to
\infty$. No discontinuity arises between the case for finite $l$ and
$l \to \infty$. Presumably, one may conclude, no energy can be
associated with the vacuum itself arises in four dimensions.

Now for the flat angular 2-manifold i.e. $k=0$ things are different.
The result (\ref{k=0}) in the asymptotically flat case $l=\infty$
studied above is twice the result we find now for finite $l$.

\begin{remark}
\label{remark} In a case $(\textbf{H})$, where no regularization is
needed for finintness,  the mass is twice the mass one finds when
the metric is looked at in the presence of an intrinsic length, as
it happens in the presence of the cosmological constant.
[It will turn out that the remark holds as long as the Einstein term
is present in the Lagrangian.]
\end{remark}

Let's see how things work out in the one dimension higher than the
usual macroscopic dimension of spacetime. The five dimensional anti
de Sitter Schwarzschild solution is a case where, as in three
dimensions, is turns out that the energy of vacuum is not zero. The
difference w.r.t. three dimensions is that vacuum energy is not
convergent for $l \to \infty$.

Let the metric be given by the five-dimensional analogue of (36)
with
\begin{equation}\label{}
g^2=\frac{r^2}{l^2}+k-\frac{r_0^2}{r^2}\,,
\end{equation}
i.e. the angular three manifold has constant curvature $k$. The
Hamiltonian will be given by
\begin{equation}\label{}
\int_{S^3}  \left(\frac{1}{2}c_1 \epsilon(i_\xi \omega E^3)+{\cal
B}_{(C)}\right)\,,
\end{equation}
for a boundary form
\begin{equation}\label{}
{\cal B}_{(C)}=\frac{1}{2}c_1 \left\{-\epsilon(\theta E^3)+b_0
\epsilon(E_\|^4)_1+b_1 \epsilon(\Omega_\| E^2_\|)_1 \right\}\,.
\end{equation}
The result is
\begin{equation}\label{}
-36\pi^2 c_1\, g^2 r^2+ 24 \pi^2 c_1 b_0\, gr^3+12 \pi^2 c_1 b_1\,
kg r \,,
\end{equation}
times $ \textit{vol}_k/(2\pi^2)$. Divergencies cancel for
\begin{equation}\label{reg-constants-5d-einstein}
b_0=\frac{3}{2l}\,, \quad  b_1=\frac{3l}{2}\,,
\end{equation}
where again $b_1$ is useless and not defined for $k=0$, and the
finite Brown-York mass is
\begin{equation}\label{5-ads-schwa}
18\pi^2 c_1 \Big(\frac{k^2 l^2}{4}+r_0^2
\Big)\frac{\textit{vol}_k}{2\pi^2} \ .
\end{equation}
For the usual definition $3! c_1=(8\pi G)^{-1}$ this agrees with the
known results in the spherically symmetric
case~\cite{Balasubramanian:1999re}\cite{Gibbons:2005jd}.

The couplings $b_0$ and $b_1$ satisfy the relation
\begin{equation}\label{renorm-couplings}
-l^2 \frac{d b_0}{dl}=\frac{d b_1}{dl}\,.
\end{equation}
We will see that when Einstein gravity is supplemented by the
Gauss-Bonnet term, the dependence of $b_0$ and $b_1$ on $l$ changes
but relation (\ref{renorm-couplings}) \emph{holds as it is}.

Clearly this result is hopeless in the limit $l \to \infty$. For
comparison purposes, consider the pure Schwarzschild metric
$g^2=1-r_0^2 r^{-2}$. This is a case (\textbf{II}). We apply
(\ref{poor-mans-hamiltonian}) and following steps explained at the
beginning of this section we find that the relative mass is given by
the limit of
\begin{equation}\label{}
-36 \pi^2 c_1\, (g-g_\alpha)g r^2+m_{r_0=0}\,,
\end{equation}
where we imply that the fields $\Psi_\infty$ are the Minkowski
solution $r_0=0$. We have
\begin{equation}\label{}
m=18\pi^2 c_1\, r_0^2+m_{r_0=0}\,.
\end{equation}
By Lorentz invariance the pure Minkowski spacetime energy should
vanish, $m_{r_0=0}=0$.

In all, formula (\ref{5-ads-schwa}) can be thought of as describing
a Schwarzschild mass given by the previous result, and a vacuum
energy given by
\begin{equation}\label{5-casimir}
m_{BY}=-\frac{3\pi}{32GK}\,,
\end{equation}
where we wrote $c_1$ in terms of the five dimensional Newton's
constant and $K=-l^{-2}$ is the constant curvature of spacetime.

We will see the result (\ref{5-casimir}) appearing again when we
consider Lovelock gravity in dimension five. It will only be
corrected by a term depending only on the gravitational couplings,
analogous to the three dimensional vacuum energy.

The vacuum energy (\ref{5-casimir}) is divergent for infinite $l$.
On the other hand the case $l=\infty$, that is asymptotically flat,
is not one such that the relative mass is a Brown-York mass i.e. is
not type $(\textbf{H})$ and the remark \ref{remark} need not apply.




\section{Point particle in three dimensions}
\label{Point particle in three dimensions}

The asymptotically flat (locally) three dimensional spherical metric
has been bypassed up to this point. It belong to the hybrid case,
type $(\textbf{H})$, and we shall apply remark \ref{remark}. We
shall discuss it at some length.

Einstein gravity in three dimensions is an always intriguing
problem. It is because it feels very elementary in some sense and to
an unimaginable extent soluble. In three dimensions there is no
curvature outside matter, i.e. no gravity in vacuum hence no
propagation of gravity through gravitational waves. In other words
the effect of gravitational sources is not local, they rather affect
spacetime in a global manner.

One implication of this is that singularities are integrable. On the
other hand all of the discussion of Hamiltonians that so emphasizes
spatial infinity is in the spirit that the gravitational effects are
diminishing there. This is not the case in three dimensions, the
deficit angle does not seize to exist for large distances.
Nevertheless things seem to work and it is interesting to apply the
formulas anyway.

Interest in Einstein gravity in three spacetime dimensions was
essentially initiated by \cite{Deser:1983tn}.

Einstein gravity in three dimensions without cosmological constant
can be described by the Lagrangian
\begin{equation}\label{}
{\cal L}=\frac{1}{2}c_1 \epsilon(\Omega E)\,,
\end{equation}
$c_1=(8\pi G)^{-1}$. The field equations say that torsion $T^A=0$
and curvature $\Omega^{AB}=0$.

The metric
\begin{equation}\label{beta-djh}
ds^2=-(dt+a \, d\phi)^2+\frac{dr^2}{\beta^2}+r^2 d\phi^2\,,
\end{equation}
is a spherically symmetric solution. $0<r<\infty$ and $0<\phi<2\pi$.
$\beta$ and $a$ are constants.

From the vielbein
\begin{equation}\label{3-veilbein}
E^0=dt+a\, d\phi\,, \quad E^1=\frac{dr}{|\beta|}\,, \quad
E^2=r\,d\phi\,,
\end{equation}
$T^A=0$ gives the connection
\begin{equation}\label{3-connection}
\omega^2_{~1}=|\beta|\, d\phi\,.
\end{equation}
Thus $\Omega=0$ and both field equations are satisfied.

Space can be given in conformally flat coordinates. Let
$r=\rho^{1-\gamma}$. Then
\begin{equation}\label{}
ds^2=-(dt+a \, d\phi)^2+\rho^{-2\gamma}(d\rho^2+\rho^2 d\phi^2)\,,
\end{equation}
where $\beta^2=(1-\gamma)^2$. The interesting thing here is the
origin $r=0$ is mapped to the origin $\rho=0$ if $\gamma<1$ and to
$\rho=\infty$ if $\gamma>1$. Also the metric is invariant under the
transformation $\rho=1/\rho'$ and $\gamma=2-\gamma'$. The other
coordinate system is not transformed: $r'=r$ and $|\beta'|=|\beta|$.

So in some sense $\rho=0$ is dual to $\rho=\infty$. There is a
difference: if $\gamma>1$, $\rho=0$ lies at infinite proper
distance, $\int_0^{\rho_0}d\rho \rho^{-\gamma}=\infty$, from any
point $\rho_0$. Instead $\rho=\infty$ lies at finite proper proper
distance from an arbitrary point, $\int_{\rho_0}^\infty d\rho
\rho^{-\gamma}<\infty$.

Let's calculate first the mass. It is the Hamiltonian for the
Killing vector is $\xi=\partial/\partial t$. Its amusing to do that
in both kinds of coordinates.

As discussed one can always treat a case as type $(\textbf{II})$ and
calculate a charge as a relative charge applying
(\ref{poor-mans-hamiltonian}).  We will do that for the moment. We
chose $\Psi_\infty$ to be the metric with $\beta^2\to 1$. The
crucial condition (\ref{matching}) is satisfied trivially here.
$T^\alpha$ is the hypersurface at some fixed radius $r_\alpha$ in
both radial coordinates we write $\Psi$ and $\Psi_\alpha$ (the need
not be the same). Using (\ref{3-veilbein}) and (\ref{3-connection})
the first line of (\ref{poor-mans-hamiltonian}) equals
\begin{equation}\label{3-hamiltonian}
\int_{S^1} c_1 \cdot \left(\frac{1}{2} \epsilon(i_\xi \omega
E)-\frac{1}{2} i_\xi \epsilon(\theta
E)\right)=-\frac{1}{4G}|\beta|\,,
\end{equation}
thus
\begin{equation}\label{3-mass}
m=\frac{1}{4G}(1-|\beta|)+m_{\beta=1}\,.
\end{equation}
The sphere at infinity is a unit circle $S^1$.
$\theta=\omega-\omega_{\|}$ is calculated for the constant radius
hypersurface $r=r_\alpha$, where $r_\alpha \to\infty$ for
$\alpha\to\infty$. This is the surface $T^\alpha$. Its
$\omega_{\|}=0$. Note that $\theta$ is the second fundamental form
only because by our derivation we have taken $T^\alpha$ to be such
that $\xi$ is tangent to it.

Consider now the conformally flat space coordinates case. The
spatial vielbein now is
\begin{equation}\label{}
E^1=\eta\rho^{-\gamma} d\rho\,, \quad E^2=\rho^{1-\gamma}d\phi\,.
\end{equation}
The connection is $\omega^2_{~1}=\eta(1-\gamma)d\phi$.

$\eta=\pm$. A hypersurface at constant $\rho$ viewed as an outer
boundary will have a normal vector in the direction of $E^1$. When
$\eta=+$ it means that we take $E^1$ in the direction of growing
$\rho$ i.e. towards infinity. When $\eta=-$, $E^1$ is in the
direction of decreasing $\rho$ i.e. towards the origin of that
coordinate.

The formula for the Hamiltonian is obtained treating $T^\alpha$ as
an outer boundary. When $\eta=+$ we integrate around the origin of
$\rho$ and when $\eta=-$ we integrate around the infinity of $\rho$.

When $\gamma<1$ the infinity of $\rho$ is at infinite proper
distance from any point. We take $\Psi_\infty$ to be the metric for
$\gamma\to 0$. Formula (\ref{poor-mans-hamiltonian}) gives
\begin{equation}\label{}
m=\frac{1}{4G}\gamma+m_{\gamma=0}\,.
\end{equation}
When $\gamma>1$ the infinity of $\rho$ lies at finite distance from
an arbitrary point $\rho_0$ while the origin is at infinite proper
distance from it: Infinity is at $\rho=0$.

We may compactify the space, add a point at infinity, and have then
two particles, one at an origin and one at infinity. Without
compactification, the `place' where a particle can be is at
$\rho=\infty$ in these coordinates.

Integrating around $\rho=\infty$ means taking $\eta=-1$. $\gamma>1$
so we cannot take $\Psi_\infty$ to the metric for $\gamma \to 0$. We
will take the `dual' choice, $\gamma \to 2$. Applying formula
(\ref{poor-mans-hamiltonian}) we find now
\begin{equation}\label{}
m=\frac{1}{4G}(2-\gamma)+m_{\gamma=2}\,.
\end{equation}

The two formulas we obtained are the two cases one obtains from
(\ref{3-mass}) from the two solutions the relations of the
parameters $\beta^2=(1-\gamma)^2$, and for the respective
`locations' one would expect form the radial transformation. This is
of course due to general covariance.

For $\gamma=1$ both $\rho=0$ and $\rho=\infty$ lie at infinite
proper distance from any $\rho_0$. It is the case of the space with
the topology of the cylinder. The candidate `places' for a particle
are the infinities of a real line. These cannot be considered as
locations of particles.

In \cite{Ashtekar:1993ds} it was argued that in the case $\gamma \ge
1$ one cannot apply uniformly the asymptotic conditions one applies
also for $\gamma<1$, because a (finite) Hamiltonian does not exist.
One may think of that as a mathematical explanation in the canonical
framework of the arguments given above.

Presumably, one may note that this discussion is hardly needed if we
use the `Schwarzschild' coordinates (\ref{beta-djh}). The
cylindrical case corresponds to $\beta=0$ which is already
unacceptable in that form of the metric.
\\

Let us now calculate the Hamiltonian generators using
(\ref{the-hamiltonian}), that is as type $(\textbf{I})$. Let's use
the minimal boundary form ${\cal
B}_{(C)}=-\frac{1}{2}c_1\epsilon(\theta E)$. The result is
essentially given by (\ref{3-hamiltonian}). We have
\begin{equation}\label{3-quasilocal-mass}
m_{BY}=-\frac{1}{4G}|\beta|\,,
\end{equation}
The result is perfectly finite, without regularization, that is we
deal with a type $(\textbf{H})$ case.

Spacetime (\ref{beta-djh}) is `spinning'. Its angular momentum
equals minus the Hamiltonian $H_\xi$ for the Killing vector
$\partial/\partial\phi$. We have:
\begin{equation}\label{}
J_{BY}=-\int_{S^1} c_1 \cdot \left(\frac{1}{2} \epsilon(i_\xi \omega
E)-\frac{1}{2} i_\xi \epsilon(\theta
E)\right)=\frac{1}{4G}|\beta|\,a\,.
\end{equation}
The term involving $\theta$ originating from the boundary does not
actually contribute, as the Killing vector is tangent onto the
sphere. In terms of the quasi-local mass (\ref{3-quasilocal-mass})
it is simply expressed as
\begin{equation}\label{3-quasilocal-spin}
J_{BY}=-m_{BY}\, a\,.
\end{equation}

There is a number of things that can be said here. Recall the
definition \ref{relative-mass-definition} for the relative
generators. Let's explicitly define a relative mass $m_\textit{rel}$
by the condition: $m_\textit{rel}=0$ for $|\beta|=1$ i.e. for
Minkowski spacetime. This fixes the free constant $H_\xi^0$. We
have:
\begin{equation}\label{relative-mass-3d}
m_\textit{rel}=\frac{1}{4G}(1-|\beta|)\,.
\end{equation}
This the value calculated in \cite{Deser:1983tn} and in any
calculation that integrates the matter energy tensor. Similarly one
may define $J_\textit{rel}$ by the condition: $J_\textit{rel}=0$ if
$a=0$. We have
\begin{equation}\label{3-spin-cry-deser}
J_{\textit{rel}}=\left(\frac{1}{4G}-m_{\textit{rel}}\right)a\,.
\end{equation}
Surprisingly relations (\ref{3-quasilocal-spin}) and
(\ref{3-spin-cry-deser}) are apparently unknown in the literature.
In Ref.~\cite{Deser:1983tn} only the case $m_\textit{rel}=0$ is
calculated, obtaining $a=-4G\, J$. The general multi-particle
massive and spinning metric was obtained in \cite{Clement:1983nk},
but the integration constants were not carefully related to
canonical generators.

Consider now the quasi-local mass $m_{BY}$ for flat spacetime. For
$|\beta|=1$ and $a=0$ we have:
\begin{equation}\label{3-flat-mass}
m_{BY}=-\frac{1}{4G}\,.
\end{equation}
In accordance with remark \ref{remark} this is indeed twice the
vacuum energy (\ref{final 3-ads mass}) for \textit{AdS} spacetime.

Reference \cite{Marolf:2006xj}, reasoning essentially along the
lines of formula (\ref{3-quasilocal-mass}), suggests that the vacuum
energy of Minkowski spacetime in three dimensions is non-zero and
equal to (\ref{3-flat-mass}). On the other hand
Ref.~\cite{Ashtekar:1993ds} using Regge-Teitelboim methods finds the
result on the r.h.s. of (\ref{relative-mass-3d}), in agreement with
\cite{Deser:1983tn}, that is the Minkowski space mass is found to be
simply zero. The results are not inconsistent: they differ by a
phase space function which an arbitrariness of the definition
anyway, relation (\ref{BY-definition}). It does raise though issues
but it is rather a matter of consistent application of definitions.
If both kinds of definitions apply they should give equivalent
physical results; if one of them is more relevant, for example in
the general setting of \textit{AdS}/CFT ideas the Brown-York
definition is the relevant one, the results should be taken
seriously whatever they are. For example Minkowski spacetime should
apparently be treated according to (\ref{3-flat-mass}).

\section{Five dimensional Chern-Simons point particle}
\label{Five dimensional Chern-Simons point particle}

We wish now to turn to the nearest analogue of the point particle in
three dimensional gravity. If analogously to three dimensions we
consider a Chern-Simons theory of the five dimensional Lorentz group
we deal with a gravity that is described by a pure Gauss-Bonnet
term. This is not exactly true unless we are in pure vacuum, but the
analogies between three and five dimensions we will find, should be
traced to the Chern-Simons equivalence of these gravities at the
level of the Lagrangian.

Presumably, working in a first order formalism, we are able to
construct Hamiltonians even in the presence of torsion. The
following are an example of this fact.

Consider the Lagrangian
\begin{equation}\label{}
{\cal L}=\frac{1}{2}c_2\, \epsilon(\Omega\Omega E)\,,
\end{equation}
where $c_2$ is a coupling constant with dimensions of mass. In
\cite{Gravanis:2007ei} it was observed that `gravity' described by
this Lagrangian admits a vacuum solution which is the obvious
generalization of (\ref{beta-djh}) for $a=0$, i.e. non spinning, and
its (relative) mass was calculated.

We consider the obvious generalization of the full spinning metric
(\ref{beta-djh}):
\begin{equation}\label{5-djh}
-(dt+a)^2+\frac{dr^2}{\beta^2}+r^2 d\Omega^2\,,
\end{equation}
where $a=a^i \tilde E^i$ for some Kerr constants $a^i$, and
$d\Omega^2=\delta_{ij} \tilde E^i \tilde E^j$ is the metric of the
unit 3-sphere, expressed conveniently in terms of a frame $\tilde
E^i$. The frame satisfies the relation $d\tilde E^i=\epsilon_{ijk}
\tilde E^j \tilde E^k$, where we may identify $\epsilon_{ijk} \equiv
\epsilon_{01ijk}$.

Now if by habit we insist that torsion is zero, then (\ref{5-djh})
does not solve the vacuum fields equations.

In Einstein gravity, when a form like $dt+a$ is introduced in the
metric one assumes $da=0$, precisely to avoid introducing torsion.
We will impose:
\begin{equation}\label{}
da=T^0\,.
\end{equation}

Under this condition the zero torsion connection of the metric with
$a^i=0$, that is $\omega^{i1}=\beta \tilde E^i$ and
$\omega^{ij}=\epsilon_{ijk} \tilde E^k$, is also a connection for
(\ref{5-djh}), but with non-trivial torsion
\begin{equation}
T^0=a^i \epsilon_{ijk} \tilde E^j \tilde E^k\,.
\end{equation}

One should note two things. First, unlike the three dimensional
Einstein case the metric (\ref{5-djh}) is not flat. The curvature
reads: $\Omega^{ij}=(1-\beta^2) \tilde E^i \tilde
E^j=(1-\beta^2)r^{-2} E^i E^j$. In fact there is a curvature
singularity at $r=0$. There is also a torsion singularity there,
$T^0_{ij}=a^i \epsilon_{ijk} r^{-2}$. The situation is somewhat
analogous: $\delta$-function singularities in curvature and torsion
in three dimensions~\cite{Jackiw:1991nb} are replaced by $r^{-2}$
singularities in five.

Second, the torsion $T^0$ needs no source. The vacuum field
equations are satisfied everywhere outside the singularity $r=0$.
Also, just like in the three dimensional Einstein gravity, the
singularity is integrable~\cite{Gravanis:2007ei}. Put differently it
is a case of type $(\textbf{H})$.

We may verify that indeed no regularization is needed. The mass is
calculated as the Hamiltonian (\ref{the-hamiltonian}) for the
Killing vector $\xi=\partial/\partial t$ from the formula
\begin{equation}\label{gauss-bonnet-hamiltonian}
\int_{S^3} c_2 \cdot \left(\epsilon(i_\xi\omega \Omega
E)-i_\xi\epsilon(\theta \big\{\Omega_{\|}+\frac{1}{3}\theta^2 \big\}
E)\right)\,.
\end{equation}
The ${\cal B}_{(C)}$ form used is the well known boundary term for
Dirichlet conditions constructed in Ref.~\cite{Myers:1987yn} (see
also the appendix). One finds easily that the quasi-local mass is
given by
\begin{equation}\label{}
m_{BY}=-8\pi^2 c_2\, |\beta| (3-\beta^2)\,.
\end{equation}
This presumably gives a Brown-York vacuum for Minkowski spacetime in
in this theory equal to
\begin{equation}\label{pure-Gauss-Bonnet-Minkowski-mass}
m_{BY}=-16\pi^2 c_2\,.
\end{equation}
It is the five dimensional analogue of the $-2\pi c_1=-(4G)^{-1}$
result of Ref.~\cite{Marolf:2006xj} for the three dimensions we
encountered in the previous section.

Defining a relative mass similarly to the three dimensions by
$m_\textit{rel}=0$ for $\beta=1$ we obtain here
\begin{equation}\label{}
m_{\textit{rel}}=8\pi^2 c_2 \left\{2-|\beta|(3-\beta^2)\right\}\,.
\end{equation}
This is the result found in \cite{Gravanis:2007ei} by integrating
the matter energy tensor. It was shown that it can be calculated
very elegantly using the Gauss-Bonnet theorem; the constant term,
which is minus the alleged Minkowski vacuum energy, is related to
the Euler number of a spatial 4-ball.

It perhaps interesting to note that, for $c_2>0$, the mass $m$ in
this example is bounded from below, with a minimum at $\beta^2=1$.
This is to be contrasted with the definite bounded-ness from above
of (\ref{relative-mass-3d}) in three dimensions, which was
especially discussed in Ref.~\cite{Ashtekar:1993ds}. In fact, the
same happens in Chern-Simons theories in all dimensions which are
multiple of four modulo one and three, respectively.

Now let $\xi^i$ be Killing vectors dual to $\tilde E^i$. The
quasi-local angular momentum is calculated as minus the
Hamiltonians, given by (\ref{gauss-bonnet-hamiltonian}), for the
vectors $\xi^i$. We have:
\begin{equation}\label{}
J^i_{BY}=8\pi^2 c_2\, |\beta|(1-\beta^2)\, a^i\,.
\end{equation}

It amusing to note the relation
\begin{equation}\label{scaling}
-3\, J^i=\beta \frac{\partial}{\partial \beta}m\, a^i\,,
\end{equation}
which also holds in three dimensions but with a factor $-1$ instead
of $-3$. Thus angular momentum is the scaling of the mass for the
Chern-Simons particles. Relation (\ref{scaling}) properly
generalizes (\ref{3-quasilocal-spin}).

\section{Boulware-Deser-Cai metrics in $5d$}
\label{Boulware-Deser-Cai metrics}

We now turn to the full Lovelock gravity in five dimensions. We
study the spherically symmetric metrics of Boulware and
Deser~\cite{Boulware:1985wk}, discovered at the same time by J.T.
Wheeler~\cite{Wheeler:1985nh}, and the generalization by
Cai~\cite{Cai:2001dz}. These are solutions to the so-called
Einstein-Gauss-Bonnet gravity in five dimensions described by the
Lagrangian
\begin{equation}\label{egb-lagrangian}
{\cal L}=\frac{1}{2}c_2\epsilon(\Omega \Omega
E)+\frac{1}{2}c_1\epsilon(\Omega
E^4)-\frac{\lambda}{5!}\epsilon(E^5)\,.
\end{equation}
The last term is formally equivalent to a `matter' energy tensor
$T_B^A=-\lambda \delta^A_B$. $\lambda$ is not an integration
constant thus it is not a phase space parameter; it operates as one
more gravitational coupling. For the Boulware-Deser-Cai solution is
a vacuum solution for this Lagrangian everywhere outside the origin
$r=0$ (which is not included in spacetime anyway). Torsion is
assumed zero which solves the connection field equations.

Abbott-Deser type of calculations were done in
\cite{Deser:2002rt}\cite{Deser:2002jk}\cite{Deser:2007vs} for
quadratic curvature gravities. The resulting formulas compute the
mass relative to the asymptotic vacuum. They are rather involved and
hard to generalize. (The mass parameter in Boulware-Deser-Cai
metrics given usually in the literature is calculated by these
formulas.)

A basic feature of Lovelock gravity is that its solutions are
multi-valued and its branches have different asymptotics. Relative
mass depends on the reference background i.e. the asymptotic vacuum.
Thus one cannot associate a value of energy with an entire solution.

Brown-York computations are in no better shape in that respect; they
too depend on asymptotics: the counter-terms depend on the branch.
On the other hand the Brown-York definition is more self-consistent:
the branch of the solution we work at is already chosen at the level
of the Lagrangian. Then a value of energy is obtained for the
solution.

Stability analysis based on energy
considerations~\cite{Deser:2002rt,Deser:2002jk,Deser:2007vs} is
treacherous in Lovelock gravity. The theory regards both branches as
available states. The Brown-York definition emphasizes strongly that
we work one branch at a time, there is no overview of the solution.
The issue of stability of vacua was touched upon in
Ref.~\cite{Charmousis:2008ce}. We shall shown in a separate work
that instabilities are far more generic in this gravity that what is
envisaged in \cite{Charmousis:2008ce}. Here we shall point out that
Brown-York masses make an unexpected appearance in semiclassical
calculations. Quasi-local calculations in Lovelock gravity have been
presented in various works, see
e.g. \cite{
Cvetic:2001bk,Padilla:2003qi,Mora:2004kb,Mora:2004rx,Dehghani:2006ws,Kofinas:2006hr,Olea:2006vd,Miskovic:2007mg,Kofinas:2007ns,Liu:2008zf,Brihaye:2008kh,Brihaye:2008xu,Brihaye:2008ns}.

The metrics can be given the general form considered in section
\ref{Spherically symmetric metrics}:
\begin{equation}\label{general-element-again}
ds^2=-g^2 dt^2+\frac{dr^2}{g^2}+r^2 d\Omega^2\,.
\end{equation}
$d\Omega^2$ is the metric of a compact, constant curvature $k$
three-manifold: denoting a frame on it by $\tilde E^i$ i.e.
$d\Omega^2=\delta_{ik} \tilde E^i \otimes\tilde E^k$ its curvature
will be $\tilde\Omega^{ij}=k\, \tilde E^i\tilde E^j$, where $k=\pm
1, 0$. The metric function $g^2$ is given by
\begin{equation}\label{egb metric 1}
g^2-k=\frac{3c_1}{2c_2}\, r^2 \Big\{1+s \sqrt{1+\frac{c_2
\lambda}{27c_1^2}+\frac{C}{r^4}}\Big\}\,,
\end{equation}
with $s^2=1$. Thus there are two solutions obtained, only one of
which can be asymptotically flat. This is one may called `Einstein'
branch. The other, $s=+1$, is the `exotic' branch of the
Boulware-Deser solution. Much of the novelty arising by these
metrics is ought to its existence.

The metrics (\ref{egb metric 1}) appear in the literature more often
in terms of the parameters $\kappa^2$, $\alpha$ and $m$:
\begin{equation}\label{}
3!c_1 \equiv \kappa^{-2}\,, \quad
\alpha\equiv\frac{c_2}{6c_1}=\kappa^2 c_2\,, \quad C=m\,
\frac{8\kappa^2 \alpha}{3\pi^2}\frac{2 \pi^2}{\textit{vol}_k}\,.
\end{equation}
The first relation implies also that $\kappa^2= 8\pi G$, in terms of
the five dimensional Newton's constant. $\textit{vol}_k$ is the
volume the 3-manifold of constant curvature $k$.
\\

Our basic formula (\ref{the-hamiltonian}) for the Hamiltonian
generator reads here:
\begin{align}\label{}
 \int_{S^3} & c_1 \cdot \left(\frac{1}{2} \epsilon(i_\xi \omega
E^3)-\frac{1}{2} i_\xi
\epsilon(\theta E^3)\right)\\
+  & c_2 \cdot \left(\epsilon(i_\xi\omega \Omega
E)-i_\xi\epsilon(\theta \big\{\Omega_{\|}+\frac{1}{3}\theta^2 \big\}
E)\right)\,, \nonumber
\end{align}
where $S^3$ is a large 3-manifold of constant curvature $k$. The
Killing vector $\xi$ is $\partial/\partial t$. The result is
\begin{align}\label{BD-mass-unreg}
 -36 \pi^2 c_1\, g^2 r^2 -24\pi^2c_2\,
g^2\big(k-\frac{g^2}{3}\big)\,,
\end{align}
times $\textit{vol}_k/(2\pi^2)$, where $\textit{vol}_k$ is the
volume of a unit 3-manifold of constant curvature $k$. This quantity
diverges for large $r$.

It will be adequate to `correct' the minimal boundary form with
terms similar to those used in the five dimensional Einstein gravity
in section \ref{Spherically symmetric metrics}:
\begin{equation}\label{}
\frac{1}{2}c_1 b_0 \epsilon(E_\|^4)_1+\frac{1}{2}c_1 b_1
\epsilon(\Omega_\| E^2_\|)_1\,.
\end{equation}
The `correction' to (\ref{BD-mass-unreg}) is
\begin{equation}\label{}
24 \pi^2 c_1 b_0\, gr^3+12 \pi^2 c_1 b_1\, kg r \,,
\end{equation}
again times $\textit{vol}_k/(2\pi^2)$.

Define the length $l$ by
\begin{equation}\label{K-EGB}
-l^{-2} \equiv K=-\frac{3c_1}{2c_2} \left\{1+s \sqrt{1+\frac{c_2
\lambda}{27 c_1^2}}\right\}\,,
\end{equation}
where $K$ is the spacetime constant curvature read off from the
asymptotics of the metric $g^2=k+r^2 l^{-2}+ {\cal O}(r^4)$. The
curvature form is $\Omega^{AB}=K E^A E^B$. This is of course
meaningful as long as $K$ of (\ref{K-EGB}) is real.

\begin{remark}
A most important fact about the structure of the counter-terms
is that they are \emph{determined solely by the asymptotic form}. In
other words: one needs only fix the constants $b_0$ and $b_1$ so
that divergencies are removed when the metric function is exactly
$g^2=k+r^2 l^{-2}$.
\end{remark}

One may consider for simplicity the \textit{AdS} case $K<0$, the
case $K>0$ can be treated analogously at least formally. One finds
that divergencies are removed for
\begin{equation}\label{fantastic-ct}
b_0=\frac{3}{2l}\left(1-\frac{2c_2}{9c_1l^2}\right)\,, \quad
b_1=\frac{3l}{2}\left(1+\frac{2c_2}{3c_1l^2}\right)\,.
\end{equation}

Then $b_0$ and $b_1$ remove the divergencies for the
Boulware-Deser-Cai metric if $l$ is given by (\ref{K-EGB}). $b_1$ is
useless and not defined for $k=0$. $b_0$ and $b_1$ make sense as
long as $l^{-1} \ne 0$ i.e. for all cases except the asymptotically
flat spacetime $s=-1$ and $\lambda=0$. On the other hand, if $k=0$
the asymptotically flat case makes sense trivially. We discuss this
interesting solution separately below. The specific counterterms
have also appeared in \cite{Brihaye:2008kh} but not in this form.

If we compare them to the couplings $b_0$ and $b_1$ we found in five
dimensional Einstein gravity in section \ref{Spherically symmetric
metrics}, the first terms are similar but there are $\alpha$
corrections. Interestingly the corrected couplings satisfy the exact
same differential relation (\ref{renorm-couplings}):
\begin{equation*}
-l^2 \frac{d b_0}{dl}=\frac{d b_1}{dl}\,.
\end{equation*}
That is, the relation depends on the same asymptotics to
\textit{AdS} and (reasonably) the dimensionality but not on the
theory.

Finally the Brown-York mass reads:
\begin{equation}\label{EGB-BY-mass}
m_{BY}=\Big(-9\pi^2 c_2+\frac{9}{2}\pi^2 c_1 l^2\Big)\,
k^2\,\frac{\textit{vol}_k}{2\pi^2}+m\,.
\end{equation}
For the spherically case $k=1$, discussed below, this formula agrees
with the result of \cite{Kofinas:2006hr} for five dimensions derived
by a set of counter-terms applying to asymptotically \textit{AdS}
spacetimes. They were constructed in the
Refs.~\cite{Mora:2004kb}\cite{Mora:2004rx}\cite{Olea:2006vd} as
boundary terms respecting the conformal symmetry of \textit{AdS}
boundary. (It was also obtained in \cite{Brihaye:2008kh} by
counter-terms similar to ours, though only for the Einstein branch
and not emphasized much). This is in accordance with our comments in
section \ref{Different boundary conditions}: If the quasi-local mass
exists i.e. the large radius is convergent, most likely the result
is independent of the boundary conditions used.

Let us write down the spherical case explicitly:
\begin{equation}\label{EGB-BY-mass-k=1}
m_{BY}=-9\pi^2 c_2-\frac{3\pi}{32 G K}+m\,.
\end{equation}
We see the mass parameter $m$, whose setting to zero makes the
metric (\ref{egb metric 1}) a metric of constant curvature $K$. We
also see a $K$ dependent term which is exactly the \emph{same} we
obtained in Einstein gravity (\ref{5-casimir}) in five dimensions.
We also se a term which depends purely in the gravitational constant
$c_2$. This is \emph{analogous} to the vacuum energy of \textit{AdS}
in three dimensions, $-(8G)^{-1}$, definitely a remnant of the
Chern-Simons nature of the Gauss-Bonnet term in five dimension
discussed in the previous section.

This result was obtained through regularization, so it is a real
type $(\textbf{I})$ case, not a type $(\textbf{H})$ one. This holds
even if we set $c_1=0$ decoupling the Einstein term, which kills the
$K$ dependent term above. Remark \ref{remark} does not apply and one
should note the difference between this first term in
(\ref{EGB-BY-mass}) and Brown-York mass of Minkowski space
(\ref{pure-Gauss-Bonnet-Minkowski-mass}) in pure Gauss-Bonnet
gravity.

The flat case $k=0$ is completely different. We noted already
writing down the counter-term coefficients (\ref{fantastic-ct}),
that in this case the limit $l \to \infty$ makes sense trivially.
(This of course happens only for Einstein branch, $s=-1$.) I.e. the
$l=\infty$ is a type $(\textbf{H})$ case. Remark \ref{remark} may
apply.

Indeed formula (\ref{EGB-BY-mass}) says that for $l<\infty$, where
regularization is still required, we have $m_{BY}=m$. But if we go
to the formula (\ref{BD-mass-unreg}) for $l=\infty$, that
$\lambda=0$ and $s=-1$, the result is convergent and equal to
\begin{equation}\label{exceptional-egb-BY-mass}
m_{BY}=2m\,.
\end{equation}
This is in accordance with remark \ref{remark} extended in Lovelock
gravity.
\\

Concluding we would like to re-calculate the mass for this metric as
a type $(\textbf{II})$ case, which we can always do, and obtain
merely a relative value for this generator. We will define it in the
usual way by: $m_{\textit{rel}}=0$ for $m=0$.

We apply formula (\ref{poor-mans-hamiltonian}) under the condition
(\ref{matching}), using the already calculated sphere integral,
formula (\ref{BD-mass-unreg}). One finds
\begin{align}
m_{\textit{rel}}= & -36 \pi^2\, c_1 \Big\{g^2 r^2- g g_\alpha r^2 \Big\}\frac{\textit{vol}_k}{2\pi^2} \\
& -24\pi^2c_2 \Big\{g^2\big(k-\frac{g^2}{3}\big)-g g_\alpha
\big(k-\frac{g_\alpha^2}{3}\big)
\Big\}\frac{\textit{vol}_k}{2\pi^2}\,, \nonumber
\end{align}
where $r=r_\alpha \to \infty$ and  $m_\alpha \to 0$ as $\alpha \to
\infty$. From that we find
\begin{equation}\label{relative-mass-egb}
m_\textit{rel}=s^2 m=m\,.
\end{equation}
The variable $m$ is the standard quantity representing the mass of
the respective spacetime in this theory. We explicitly saw here that
this follows from a subtraction method of calculation. There are two
things that should be remarked upon.

First, the previous result holds except for $k=0$ and $\lambda=0$ of
the Einstein branch ($s=-1$) of the Boulware-Deser-Cai metric, which
is a type $(\textbf{H})$ metric as we already mentioned. In that
case one finds
\begin{equation}
m_\textit{rel}=2m\,.
\end{equation}
Comparing with (\ref{exceptional-egb-BY-mass}) we see that they are
equal. This is in accordance with the very definition of a type
$(\textbf{H})$ metric: the relative mass can be regarded as a
Brown-York mass.

Secondly, there is a new thing brought up by considering relative
values of generators in Lovelock gravity. All metrics are at least
double-valued in this theory, thus so do the vacuum metrics. The
relative mass in each branch is calculated with respect to the
constant curvature spacetime (vacuum) in \emph{that} branch. As the
vacua are very different so is the meaning of parameter $m$ as the
mass in each branch. Put in different words, the masses in each case
are defined for different asymptotics and therefore their values are
not comparable. For $\lambda=0$ this is all we have: the Einstein
branch ($s=-1$) is asymptotically flat.

When the cosmological constant is non-zero the Brown-York definition
of the generators raises this fake degeneracy in an explicit though
formal manner. The Brown-York masses of the two branches of a given
Boulware-Deser metric differ by
\begin{equation}\label{BY-EGB-difference}
m^\textit{Einstein}_{BY}-m^{exotic}_{BY}=-162\pi^2\,
\frac{c_1^2}{\lambda}\,\sqrt{1+\frac{c_2 \lambda}{27 c_1^2}}\,.
\end{equation}

Of course this subtraction is meaningful for comparison purposes
when we subtract things of the same nature. On the other hand
$m_{BY}$ is obtained by adding to the action counterterms which
depend on the scale $l$. $l$ in turn, defined in (\ref{K-EGB}),
depends explicitly on $\lambda$ as well on the branch sign $s$,
distinguishing the branches already at the level of the Lagrangian.
This is like comparing masses of completely different gravitational
theories. Therefore one cannot regard the Brown-York mass as a kind
of absolute mass and cannot regard the $m_{BY}$ in each branch as
comparable quantities. That is, also (\ref{BY-EGB-difference}) is an
formal subtraction.

On the other hand if we restrict ourselves to the sector of
solutions belonging to a given branch then the Brown-York mass can
be regarded as a kind of absolute definition in the following sense:
One calculates the mass a Hamiltonian generator of a given theory
with couplings $c_1$, $c_2$ and $\lambda$, and no reference
background or subtractions are involved. Then, inspection of the
formulas (\ref{EGB-BY-mass-k=1}) and (\ref{K-EGB}) show that the
Brown-York mass increases with $\lambda$ in the Einstein branch
sector of solutions and decreases with $\lambda$ in the exotic
branch sector. This is a reflection in the quasi-local calculation
of the anti-gravity behavior of exotic branch solutions:
asymptotically the exotic branch metric has an extra minus sign in
the mass-dependent term.

As a physical energy of some kind, Brown-York mass is inherently
related to quantum effects through the \textit{AdS}/CFT
correspondence~\cite{Maldacena:1997re}\cite{Witten:1998qj}\cite{Gubser:1998bc}:
$m_{BY}$ equals the vacuum energy (Casimir effect) of a conformal
quantum field theory living on the boundary manifold. This is a
purely quantum effect coming from the zero-point energies of the
field oscillators. We shall not go into boundary dual field theories
here but we can point to another quantum place where the Brown-York
mass makes an appearance.

In the semiclassical description of false vacuum decay (in the
presence also of
gravity)~\cite{Coleman:1977py}\cite{Callan:1977pt}\cite{Coleman:1980aw},
one works in imaginary (Euclidean) time and considers configurations
such that a bubble of real vacuum nucleates within a sea of false
vacuum. The quantity which gives the rate of nucleation involves the
on-shell Euclidean action. It is made of pieces of the form:
$\textrm{(energy density) $\times$ (bubble volume)}$, and
$\textrm{(surface tension) $\times$ (bubble area)}$. For transitions
between vacua of some curvatures $K$ one may verify that the bulk
pieces of the on-shell action are proportional to
\begin{equation}\label{}
m_{BY}\cdot K^2\cdot \textrm{(bubble volume)}\,,
\end{equation}
where $m_{BY}$ is explicitly given by (\ref{EGB-BY-mass-k=1}) for
$m=0$. From the analogous formulas in Einstein gravity it is not
very clear that the Brown-York mass appeared there, or better a
quantity equal to it, as the result is what one expects on
dimensional grounds. In Lovelock gravity there are relative
coefficients that should match. We do not have a detailed
explanation for this phenomenon.

\section{Boulware-Deser metrics in $6d$}
\label{Boulware-Deser metrics in $6d$}

Dimension five, being the minimum dimension for the quadratic
Lovelock term to exist, exhibits certain peculiarities inexistent in
dimension six or higher. This is shown for example in the
qualitative differences of the black hole solutions between five and
higher dimensions~\cite{Cai:2001dz}, or in the novel topological
black holes presented in \cite{Bogdanos:2009pc} which exist in
dimension six and not in five due to a vanishing Weyl tensor. Five
dimensions being minimal could be treacherous. The simple gauge
theoretic construction of counter-terms we advocate could fail in
dimension higher than five. This is not so. To our knowledge such a
calculation has been not been presented in the literature.

The Lagrangian of the theory is easily written down, according to
what we have said already in the introduction. It amounts to an
insertion of one more factor of $E$ in the Lagrangian
(\ref{egb-lagrangian}) (and divide $\lambda$ in the last term with
6! instead of 5!). The field equations are obtained by varying the
action functional with respect to $E$. For the standard form of the
metric (\ref{general-element-again}) the field equations give
\begin{equation}\label{egb-in-6d}
g-k=\frac{c_1 r^2}{c_2}\Big\{1+ s \sqrt{1+\frac{c_2\lambda}{120
c_1^2}+\frac{C}{r^5}}\Big\}\,.
\end{equation}
$C$ is an integration constant associated with the mass of the
spacetime w.r.t. the asymptotic vacuum. Of course there are two
branches in the solution, $s=\pm$. The relation of our couplings
$c_1$ and $c_2$ to the usual couplings is: $4!c_1=\kappa^{-2}$ and
$\alpha=2!\kappa^2 c_2$, where $\kappa^2=8\pi G$.

The Brown-York Hamiltonian (\ref{the-hamiltonian}), constructed out
of the minimal Dirichlet boundary form, reads
\begin{align}\label{}
 \int_{S^4} & c_1 \cdot \left(\frac{1}{2} \epsilon(i_\xi \omega
E^4)-\frac{1}{2} i_\xi
\epsilon(\theta E^4)\right)\\
+  & c_2 \cdot \left(\epsilon(i_\xi\omega \Omega
E^2)-i_\xi\epsilon(\theta \big\{\Omega_{\|}+\frac{1}{3}\theta^2
\big\} E^2)\right)\,, \nonumber
\end{align}
where $S^4$ is a large 4-manifold of constant curvature $k=\pm 1,0$.

This diverges so we supplement the boundary form ${\cal B}_{(C)}$
with the intrinsic counter-terms
\begin{equation}\label{}
\frac{1}{2}c_1 b_0\, \epsilon(E_\|^5)_1+\frac{1}{2}c_1 b_1\,
\epsilon(\Omega_\| E_\|^3)_1+\frac{1}{2}c_1 b_2\,
\epsilon(\Omega_\|\Omega_\| E_\|)_1\,.
\end{equation}
One finds
\begin{align}
&-256\pi^2\, c_1\, g^2 r^3-256\pi^2 c_2\, g^2\Big(k-\frac{g^2}{3}\Big)r \\
&+32\pi^2\, c_1b_0\, 5 g r^4+32\pi^2\, c_1b_1\, 3k g r^2+32\pi^2\,
c_1b_2\, k^2 g\,, \nonumber
\end{align}
times $ 3\,\textit{vol}_k/(8\pi^2)$, where $\textit{vol}_k$ is the
volume of the constant curvature $k$ four-manifold $S^4$. This
additional factor is equal to one when $S^4$ is a sphere.

We may consider for simplicity asymptotically \textit{AdS} (or flat)
metrics. The asymptotic \textit{AdS} length scale $l$  and constant
curvature $K$ are defined by
\begin{equation}\label{l-for-6d}
-l^{-2} \equiv K=-\frac{c_1}{c_2}\Big\{1+ s
\sqrt{1+\frac{c_2\lambda}{120 c_1^2}}\Big\}\,.
\end{equation}

We have already mention that removing the divergencies for the
asymptotic metric alone, that is when the metric is exactly given by
$g^2=k+r^2 l^{-2}$, suffices to remove the divergencies for all
metrics (\ref{egb-in-6d}). It is then straightforward to verify that
the divergencies are removed when
\begin{align}
& b_0=\frac{8}{5l}\left(1-\frac{c_2}{3c_1 l^2}\right),\\
& b_1=\frac{4l}{3}\left(1+\frac{c_2}{c_1 l^2}\right), \nonumber \\
& b_2=-l^3\left(1-\frac{3c_2}{c_1 l^2}\right). \nonumber
\end{align}
$b_1$ and $b_2$ are useless and not defined for $k=0$.

Let now $l$ be given by (\ref{l-for-6d}). The Brown-York mass for
the spacetime with metric (\ref{egb-in-6d}) in the respective branch
reads
\begin{equation}\label{}
m_{BY}=C\frac{c_1^2}{c_2}s^2=C\frac{c_1^2}{c_2}\,,
\end{equation}
times $3\,\textit{vol}_k/(8\pi^2)$. $s=\pm$ is the branch sign. This
result is in agreement with quasi-local calculations in the
literature~\cite{Kofinas:2006hr}\cite{Brihaye:2008ns}\cite{Liu:2008zf}.
We derived the result from scratch in a relatively easier manner.

It is worth to mention the following. The Brown-York mass of pure
\textit{AdS}, or for that matter of pure de Sitter spacetime, given
by $C=0$, is zero in six dimensions. The same we observed in four
dimensions, and holds for all even
dimensions~\cite{Aros:1999kt}\cite{Kofinas:2006hr}. As a result the
Brown-York of both branches of the solution is given by a similar
formula.

\begin{center} *** \end{center}

From the above we reach the conclusion: If we have a Lovelock
gravity in the bulk, all is required to have is a Lovelock gravity
intrinsic to the boundary. Then all divergencies can be removed for
a pure \textit{AdS} background by appropriately fixing the couplings
of the boundary Lovelock theory. Then the same boundary theory
applies also to asymptotically \textit{AdS} spacetimes. Formally, it
should not be very difficult to determine the $b$ coefficients in
the general Lovelock gravity, or even to more general geometric
theories. A better understanding of the whole thing would be much
better and we shall leave such an analysis for a separate work.

\section{Flat from \textit{AdS}}
\label{interpolations}

The quasi-local charges $H_\xi(\alpha)$, relation
(\ref{the-hamiltonian-quasilocal}), were introduced as candidates of
the total content of a spacelike section $\Sigma^\alpha$ bounded by
$S^\alpha$ in the respective conserved charge. Such a quantity
should possess a number of properties, discusses in the works cited
in section \ref{conservation}. Of them we prove none apart from the
naturalness of their definition i.e. them being essentially on-shell
values of the Hamilton-Jacobi theory generators. On this basis at
least, one may study them and realize that one can extract
interesting information from them. For example they operate as
interpolations between the singular asymptotical flat case and the
non-zero cosmological constant metrics.

As a first example consider \textit{AdS} metric in three dimension,
discussed in section \ref{Spherically symmetric metrics}. The mass
was calculated there as the large $r$ limit of the quantity $-2\pi
c_1\, g^2-2\pi c_1\, b_0\, gr$. After a little algebra one may write
the result in the form
\begin{equation}\label{m(r)-ads-3}
m_{BY}(r)=-\frac{1}{4G}\bigg(1+\frac{1}{\sqrt{1+(l/r)^2}}\bigg)^{\!-1}\,.
\end{equation}
One first observes that this is an increasing function of $r$: it is
larger the larger is the size of the region in space. Specifically
goes from the value $-(4G)^{-1}$ obtained in the `small sphere'
limit $r \to 0$, to the value $-(8G)^{-1}$ in the limit $r \to
\infty$. The latter value is of course the value of the Brown-York
mass (\ref{final 3-ads mass}) of \textit{AdS} spacetime in dimension
three.

On the other hand, the `small sphere' result $-(4G)^{-1}$ can be
recognized as the Minkowski spacetime Brown-York mass
(\ref{3-flat-mass}) we found in section \ref{Point particle in three
dimensions}. The reason is simple: The result depends on the lengths
$l$ and $r$ only through the dimensionless ratio $l/r$. Therefore $r
\to 0$ is equivalent to $l \to \infty$. This is nothing but taking
the asymptotically flat limit before letting the size of the region
to go to infinity. Letting $r$ become large is then trivial in this
example and one obtains from the \textit{AdS} space quasilocal
charge (\ref{m(r)-ads-3}) the flat space Brown-York mass.

Thus there is actually continuity in the limit $l\to \infty$. (This
is also one more reason why we should apparently take a non-zero
Brown-York defined mass for the Minkowski spacetime seriously). The
reason why it is possible obtain the flat space result for the
\textit{AdS} space one, is that the former is a type $(\textbf{H})$
case.
One may verify that everything said above applies also to the
five-dimensional $k=0$ Boulware-Deser-Cai metric of Lovelock
gravity.

We may now extend remark \ref{remark} which concerns the types
$(\textbf{H})$ cases:
\begin{remark}
The quasi-local values of the Hamiltonian generators in type
$(\textbf{H})$ cases can be obtained as the `small sphere' limit of
the quasi-local generators of the associated \textit{AdS} metrics.
\end{remark}
The need for the extension is that in the limit $l \to \infty$ the
mass not always changes by a factor of 2. This can be seen by
comparing the mass (\ref{EGB-BY-mass-k=1}) of the Boulware-Deser
metric and the mass (\ref{pure-Gauss-Bonnet-Minkowski-mass}) of the
Minkowski spacetime in pure Gauss-Bonnet gravity (i.e. no Einstein
term is present) which is a type $(\textbf{H})$ case.

To make things completely explicit let's consider pure Gauss-Bonnet
gravity in the presence of a cosmological constant. This is the
theory (\ref{egb-lagrangian}) for $c_1=0$. Everything we need can be
obtained from the full Lovelock gravity results of the previous
section, as neither the length $l$ nor the counter-term coefficients
$c_1 b_0$ and $c_1 b_1$ are singular in setting $c_1=0$. In any case
it is straightforward to verify that the obtained counter-terms
work. One finally finds
\begin{align}\label{}
&m_{BY}(r)=4\pi^2c_2\times \\
&\times\Bigg\{\bigg(3\frac{r^2}{l^2}-2\frac{r^4}{l^4}\bigg)\bigg(\sqrt{1+\frac{l^2}{r^2}}-1\bigg)+\frac{r^2}{l^2}-4\Bigg\}\,.
\nonumber
\end{align}
In the limit $r \to \infty$ we obtain $-9\pi^2c_2$. This is the
first term of the full formula (\ref{EGB-BY-mass-k=1}) as expected.
In the small sphere limit $r \to 0$ one obtains $-16 \pi^2 c_2$
which indeed is the $l=\infty$ result
(\ref{pure-Gauss-Bonnet-Minkowski-mass}).

Now, if $m_{BY}(r)$ is the amount of total energy contained in a
sphere of radius $r$ in these examples, why does it not vanish in
the small sphere limit?

First of all one can always set a constant value of energy to zero
by suitably fixing the constant $H^0$ in (\ref{the-hamiltonian}) and
(\ref{the-hamiltonian-quasilocal}). Consider then the example of
\textit{AdS}$_3$. If we fix $H^0$ such that $m_{BY}(r)$ vanishes in
the small sphere limit then the Brown-York mass in spacetime is
$-(4G)^{-1}$, instead of $-(8G)^{-1}$. But it is not that easy to
change our minds and fiddle the latter value; it can be
re-discovered as the mass of a state in the boundary conformal field
theory which can be identified with
\textit{AdS}$_3$~\cite{Strominger:1997eq}. It is then advisable to
relax the interpretation of the quasi-local mass as the mass
`contained' in a spatial region. A better use of it has been pointed
out in this section.

\section{summary and comments}

Covariant phase space methods are used in the first order
formulation of Einstein and Lovelock gravity for the derivation
of the Hamiltonian generators. Many known results and some new ones
are derived and discussed in this context. Relations between them
and phenomena which arise as we go from three to six dimensions are
studied. We find that in the odd dimensions, if the higher possible
Lovelock term is included, one should attribute a Brown-York energy
in Minkowski spacetime. This is intimately related to the effect of
the latter to create deficit angles singularities in space. The
relation of the flat spacetime results to those of the associated
\textit{AdS} spaces is discussed. In certain cases, where a minimal
boundary Lagrangian is adequate for convergence, the flat space
result derives from the small sphere limit of the \textit{AdS} one.
In higher dimensions, five and six, with Lovelock gravity turned on,
the simplicity of the first order formulation of the theory (in
differential forms notation) with Dirichlet boundary conditions and
its agreement with other methods is emphasized. This agreement is
consistent with the analysis of the first sections of the paper
which suggests that all convergent quasi-local definitions should
agree up to a phase space constant. The counter-term coefficients
exhibit a structure: for example, in five dimensions those
coefficients of the full Lovelock gravity obey the same differential
relation with those of Einstein; something similar can be done in
dimension six. It becomes rather clear that a boundary Lovelock
gravity can regularize the divergencies of a Lovelock gravity in the
bulk. Study of this general problem is left for future work.

Well defined Hamiltonian generators rest on the existence of a
non-degenerate symplectic form, which we tacitly assumed. This is
not guarantied in Lovelock gravity. In an analogous particle system
with Lagrangian $L(x,v)$ the symplectic form would read
$d(dL/dv)\!\wedge\!dx$. This gives trouble if $L$ has an inflection
point. At such points accelaration terms drop out the field
equations. Neither such a system nor Lovelock gravity have an priori
well posed initial value problem~\cite{ChoquetBruhat:1988dw}. We may
derive a Hamiltonian in the space of solutions only away from those
points~\cite{TZ}.

An interesting way out of this problem, as well as out of worst
problems of causal evolution arising for piecewise smooth
solutions~\cite{Garraffo:2007fi}, was put forward in
Ref.~\cite{Willison:2009fc}. There is a class of Lovelock gravities,
which are not too special i.e. they are not of measure zero in
coupling space, whose actions are interpreted as a Weyl tube volume
(see \cite{Willison:2009fc} and references therein). The idea is
that as long as the Weyl tube does not intersect itself, in which
case a Lovelock action does represent the tube volume, the
associated gravity is safe. To the extent this is correct one could
consistently do Hamiltonian theory in those theories.

\acknowledgments We thank S. Willison for helpful discussions. Part
of this work was done during a pleasant visit to CECS (Centro de
Estudios Cient\'{\i}ficos); the Center is thanked for hospitality.

\section*{Appendix}

\appendix

\numberwithin{equation}{section}

\section{Second fundamental form}
\label{theta}

Consider a non-null hypersurface embedded in spacetime with a unit
vector field $\zeta^A$ normal to it. $\zeta \cdot \zeta=\pm 1$. We
denote by $i^*_{\|}$ the pull-back into the hypersurface. Let
$E^a_{\|}$ be an intrinsic vielbein and $\omega^{ab}_{\|}$ be an
intrinsic connection. Lower case Lorentz indices label tensor
components normal to $\zeta$.

Impose a coincidence condition:
$i^*_{\|}\omega^{ab}=\omega_{\|}^{ab}$ and $i^*_{\|}
E^{a}=E_{\|}^{a}$ i.e. the induced fields coincide with the
intrinsic fields. This in accordance with section \ref{action and
symplectic form}. This is possible in each smooth component of the
boundary.

Let us impose the conditions that the induced fields held fixed:
\begin{align}
\label{no-variation-of-induced} & i^*_{\|}\delta \omega^{ab}=0 \quad
\textrm{and} \quad i^*_{\|} \delta E^a=0\,.
\end{align}
Then also $\delta E^a_{\|}=0$ and $\delta \omega^{ab}_{\|}=0$, by
the coincidence condition. In words: the induced fields are not
varied under Euler-Lagrange variations; respecting the coincidence
condition the intrinsic fields are held fixed to their values. This
is the `Dirichlet boundary conditions' in first order formalism.

We now define
\begin{equation}\label{theta_definition}
\boldsymbol\theta:=i^*_{\|}(\omega-\omega_{\|})\,.
\end{equation}
This is the second fundamental form of the embedding of the
hypersurface into spacetime. Clearly, the non-zero components of
$\boldsymbol{\theta}$ have one index in the normal direction of the
hypersurface.

The fields $\omega_{\|}$ and $E_{\|}$ can be regarded as a
\emph{bulk} fields which agree with the intrinsic fields when pulled
back into the hypersurface. We often use the quantity $\theta \equiv
\omega-\omega_{\|}$. Of course $\boldsymbol\theta=i^*_{\|} \theta$.

\section{boundary forms}
\label{boundary forms}

Consider for simplicity Einstein gravity in three dimensions. The
Lagrangian is ${\cal L}=\frac{1}{2}c_1\epsilon(\Omega E)$.

We have
\begin{equation}\label{simple_variation}
\delta \epsilon(\Omega E)=d\epsilon(\delta \omega E)+\delta \Psi
\cdot \mathcal{E}\,.
\end{equation}
We have
\begin{align}\label{dirichlet_bdy}
i^*_{\|} \epsilon(\delta \omega E) & =i^*_{\|}\epsilon\left(\delta[\omega-\omega_{\|}]E\right)+i^*_{\|}\epsilon(\delta\omega_{\|}E)\\
&=\delta\{i^*_{\|}
\epsilon\left([\omega-\omega_{\|}]E\right)\}+i^*_{\|}\delta \Psi
\cdot \mathcal{E}_{\textit{bdy}}\,,\nonumber
\end{align}
where $\mathcal{E}_{\textit{bdy}}$ are the boundary equations of
motion, essentially the quasi-local energy momentum and spin tensors
$Q_{\textit{quasi-local}}$ of section \ref{action and symplectic
form}. By (\ref{no-variation-of-induced}) this term vanishes.

Thus we have that under Dirichlet conditions
\begin{equation}\label{}
\delta_{(C)}\left\{\int_{\cR} \epsilon(\Omega E)-\int_{\bcR}
\epsilon\left([\omega-\omega_{\|}]E\right) \right\}=\int_{\cR}
\delta_{(C)} \Psi \cdot \mathcal{E}\,.
\end{equation}
The symbol $\delta_{(C)}$, which we use in the main text, emphasizes
that the variations are done under the specific boundary conditions.
Therefore the action
\begin{equation}\label{}
S=\frac{1}{2}c_1\int_{\cR} \epsilon(\Omega
E)-\frac{1}{2}c_1\int_{\bcR} \epsilon\left(\theta E\right)\,,
\end{equation}
has an extremum on-shell under $\delta_{(C)}$. This is
Einstein-Hilbert action with a Gibbons-Hawking term. The boundary
form $\mathcal{B}_{(C)}$ used in the next is read off from it. The
generalization of this action in higher dimensions amounts to simply
inserting an equal number of factors of $E$ in the bulk and boundary
form, as one may verify.

In Einstein-Gauss-Bonnet gravity, we mostly use in this paper, the
boundary form was written down in Ref.~\cite{Myers:1987yn}, and for
Lovelock gravity in Ref.~\cite{MuellerHoissen:1989yv}. This can be
constructed in all Lovelock gravities via a Chern-Weil procedure
which provides the transgression forms in the associated topological
problems, see e.g.~\cite{Eguchi:1980jx}. In that spirit they where
constructed in \cite{Gravanis:2003aq}. A general method for
construction of the boundary forms for general geometric Lagrangians
was presented in \cite{Gravanis:2009pz}.

\bibliography{EnergyFormula_published}      
\bibliographystyle{h-physrev5}
\end{document}